\documentclass[10pt,letterpaper]{article}
\usepackage{opex3}
\usepackage{latexsym}

\begin{document}
\title{Customised broadband metamaterial absorbers for arbitrary polarisation}

\author{Hiroki Wakatsuchi$^*$, Stephen Greedy,\\ Christos Christopoulos$^*$, and John Paul}

\address{George Green Institute for Electromagnetics Research, School of Electrical and Electronic Engineering, University of Nottingham, Tower Building, University Park, Nottingham, \\NG7 2RD, U.K.}

\email{hirokiwaka@gmail.com, Christos.Christopoulos@nottingham.ac.uk} 



\begin{abstract}
%
This paper shows that customised broadband absorption of electromagnetic waves having arbitrary polarisation is possible by use of lossy cut--wire (CW) metamaterials. These useful features are confirmed by numerical simulations in which different lengths of CW pairs are combined as one periodic metamaterial unit and placed near to a perfect electric conductor (PEC). So far metamaterial absorbers have exhibited some interesting features, which are not available from conventional absorbers, e.g. straightforward adjustment of electromagnetic properties and size reduction. The paper shows how with proper design a broad range of absorber characteristics may be obtained. 
\end{abstract}



%

\section{Introduction}
Recently the use of metamaterials as wave absorbers has been attracting the interest of many researchers from the viewpoint of feasible metamaterial applications using current technology. Since the first experimental metamaterial work was reported by Smith et al. \cite{smith}, various metamaterials have been proposed to realise their exotic properties, such as negative permittivity, negative permeability and negative refractive index (NRI), from the microwave region to the optical region \cite{caloz,physicasReports,SoukoulisMTMs}. At the same time many metamaterial applications have also been suggested; the superlens \cite{superlens}, invisible cloaking \cite{opticalTransformation,scatteringCancel} and optical transformation \cite{optTrans} to name but a few. However, there are obstacles in the realisation of these schemes. Most metamaterial applications rely on the negative properties described above. At high frequencies, such as terahertz and optical bands, the properties have non--zero imaginary parts of refractive index and wave impedance mainly due to the conductive loss of the metal components, making the applications less attractive or sometimes unrealistic. For this reason the conductive loss issue is recognised as a problem to be solved \cite{scienceColumnNextMTM} and many researchers are trying to address this matter, for example, by the use of gain mediums \cite{gainMTM} or by optical transformation designs that include the conductive loss \cite{superscatterer}. In this paper we seek to exploit these losses for efficient absorber design. In metamaterial absorbers the use of the conductive loss can be taken into account as part of the absorber design to achieve strong levels of absorptance. Thus the practical use of metamaterials as wave absorbers is becoming popular for applications in areas such as EMC, crosstalk reduction etc. 

So far metamaterial absorbers have exhibited attractive features which are not available from conventional wave absorbers. One example is in the manipulation of the electromagnetic properties. In conventional absorbers design options are limited by substrate properties and mixture ratios to realise desired permittivity and permeability values. This is not easy to accomplish. In this paper we aim at systematic design techniques for metamaterial absorbers to enable us to modify their electromagnetic properties with relative ease (e.g. changes in metallisation length and geometry). More importantly the use of metamaterials makes the absorber substrate thicknesses significantly smaller. In the conventional absorbers the total thickness of the structures need to be comparable to a quarter wavelength $\lambda _0$ of the operating frequency. However, the $\lambda _0/4$ thickness can limit the range of application, for example, where available working space is restricted. This issue has successfully been solved by use of thin metamaterials, which allowed us to relax the limitation of the physical size remarkably (e.g. in \cite{ultraThinAbs} 1 mm thickness at operating frequency 6 GHz, i.e. approximately one fiftieth of the wavelength), and more flexible design options became possible. 

An outstanding issue is the limitation of the absorption band width. In most cases absorption is concentrated in a narrow frequency band. Although narrow band absorbers may be useful in some applications, the real need is for broadband absorbers with customised characteristics to meet specific user needs. 

This paper shows that highly customisable broadband absorption is possible using metamaterials as wave absorber. Different lengths of conductive and lossy cut--wires (CW) (see insets of Figs. \ref{fig:modelField} (a) and (b) as examples of CWs) are deployed as a single periodic metamaterial absorber unit. Highly customisable absorption is achieved by independent conductive loss optimisation in each CW. In addition, this absorption can be obtained for arbitrary polarisations without any serious absorptance reduction when other pairs of CWs are introduced. Furthermore, it is also demonstrated that not only enhancing the absorptance but also reducing it is possible by use of additional CWs. These results indicate that the CWs make it possible to design very flexible absorptance characteristics. Before the customisable absorber configurations are explained in detail, this paper introduces characteristics of lossless CW metamaterials in section \ref{sec:material} then examines the basic properties of CW metamaterials in sections \ref{sec:singleAndPaired}. In section \ref{sec:CWpec} the customised broadband absorber designs are introduced and conclusion is made in section \ref{sec:conc}. Previous work is described in \cite{APRASC} which was done using limited numerical resolution. This work is extended here to develop customised absorbers. 

\section{Calculation models}
\label{sec:material}
This paper uses a thin rectangular metal form, the so--called CW, as a base metal geometry for the following reasons. Firstly, the CW metamaterials are very simple structures and the periodic unit is composed of only a dielectric substrate and a single or paired CW. Due to their simplicity, CW metamaterials can be transformed into other types of metamaterials \cite{CWeq,CWfromSRR}, so they play an important role in basic metamaterial research. Secondly, the CW metamaterials do not require any extra metal components to yield a NRI. For example, the first metamaterial \cite{smith} is composed of split--ring resonators, yielding a negative value of permeability, and long metal strips, yielding a negative value of permittivity, and hence two types of metal structures are necessary to produce a NRI. However, the CW metamaterials can exhibit both negative permittivity and permeability through a simply paired CW. Thirdly, CW structures are suitable for introducing stronger losses. As is reported in \cite{SoukoulisMTMs}, compared to fishnet structures \cite{fishnetKRH} which have some similar properties with those CW metamaterials have, the CW metamaterials can exhibit greater losses, which although undesirable in most metamaterial applications they are needed in absorber applications. Moreover, from a fabrication viewpoint, the CW form is straightforward to manufacture due to its simple geometry. Finally, CWs with conductive loss easily produce multi--absorptance peaks as will be demonstrated in this paper. 

In absorber applications described in this paper conductive loss is used instead of dielectric loss which is assumed to be a primary resource in most metamaterial absorbers (e.g. \cite{dieLossAbs}). One of the reasons is that materials with suitable conductive losses are readily available through recently developed technologies, e.g. conductive inks \cite{conductiveInk}. Also, independent optimisation of the metal resistances gives us more flexible options for absorption characteristic designs, e.g. multi--peak broadband absorption, as will be demonstrated in section \ref{sec:CWpec}. 

\begin{figure}[htbp]
\begin{minipage}{0.5\hsize}
\centering
\includegraphics[width=\linewidth]{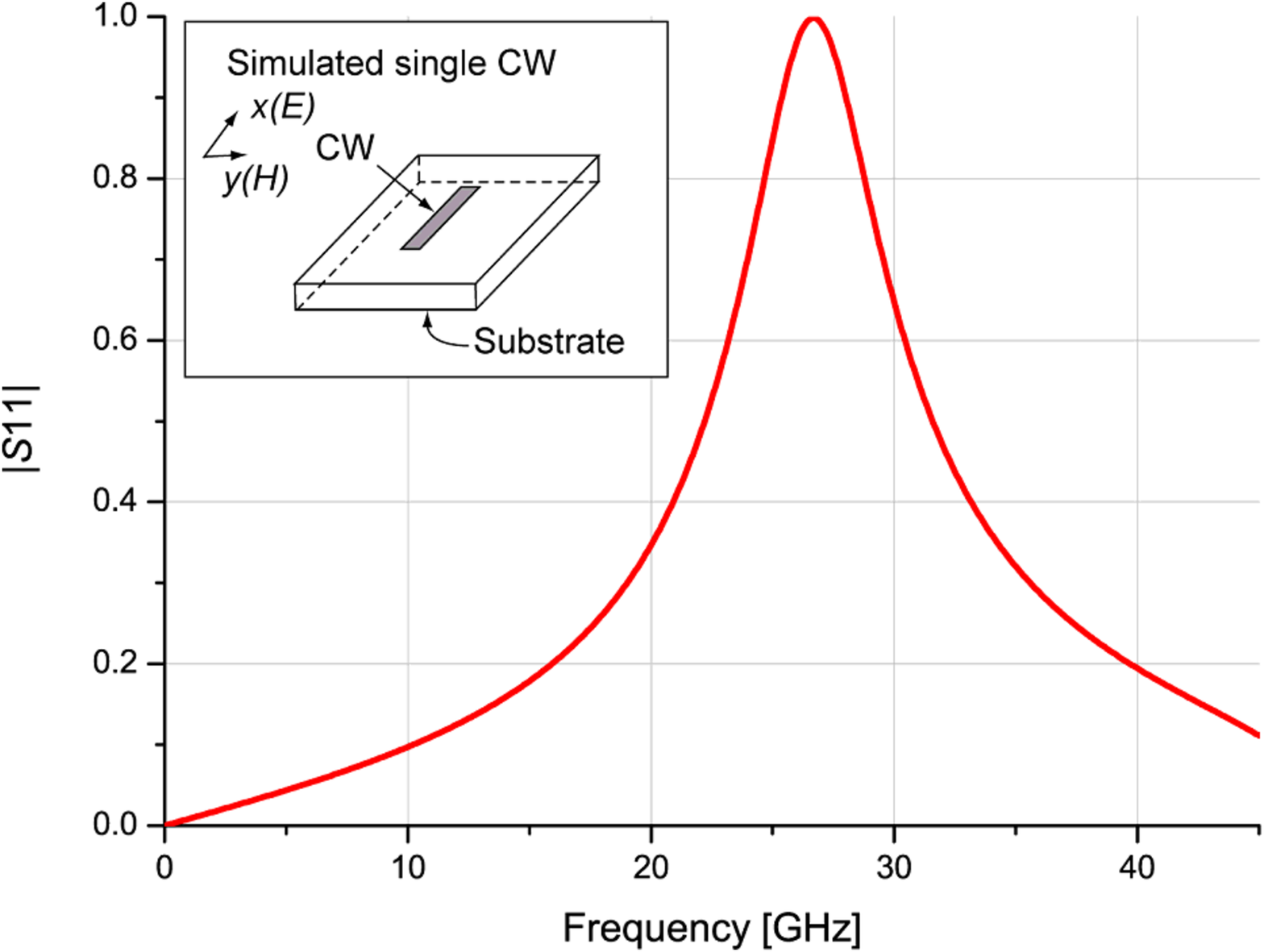}
\end{minipage}
\begin{minipage}{0.5\hsize}
\centering
\includegraphics[width=\linewidth]{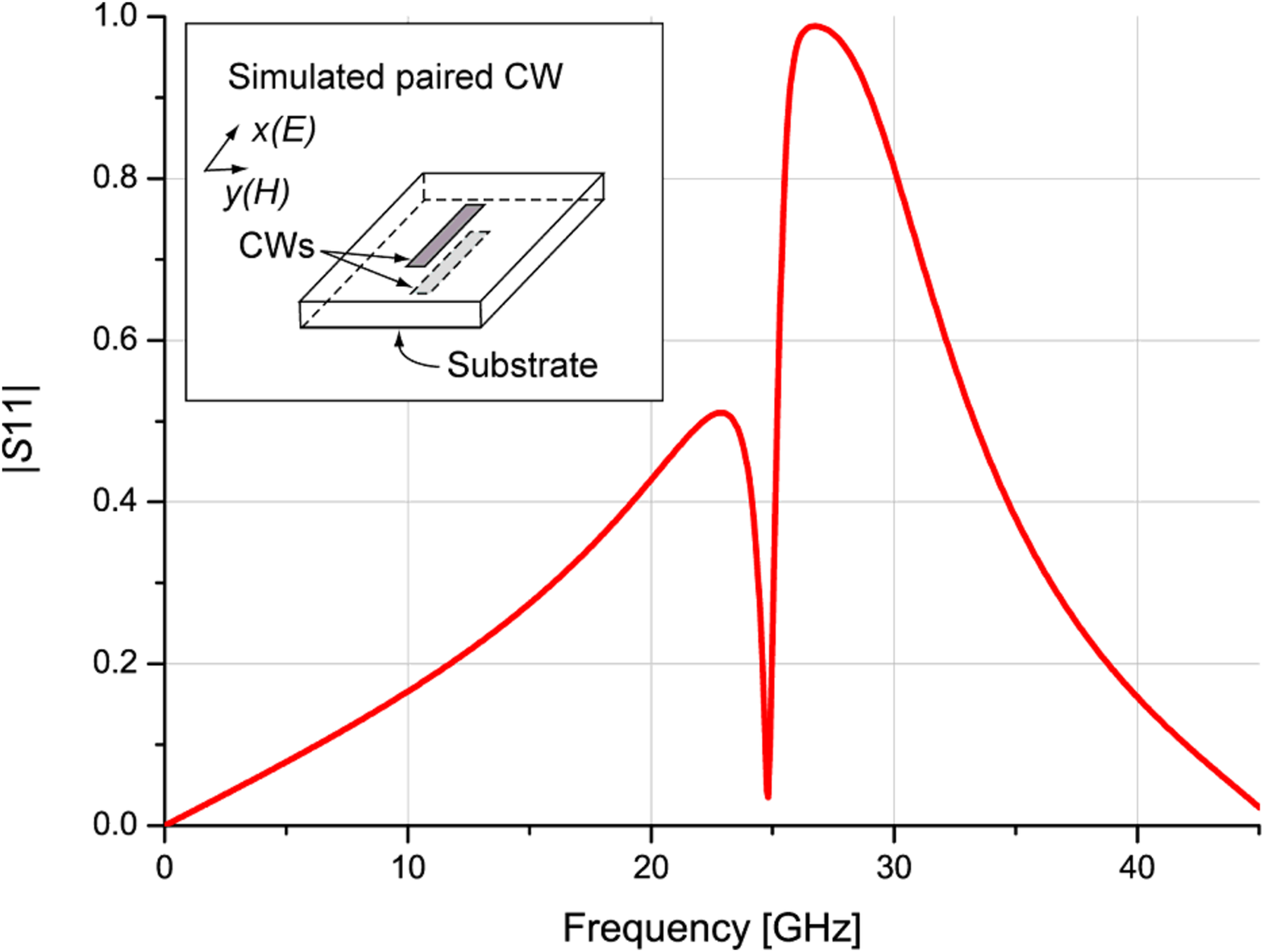}
\end{minipage}
\begin{minipage}{0.5\hsize}
\centering
\includegraphics[width=\linewidth]{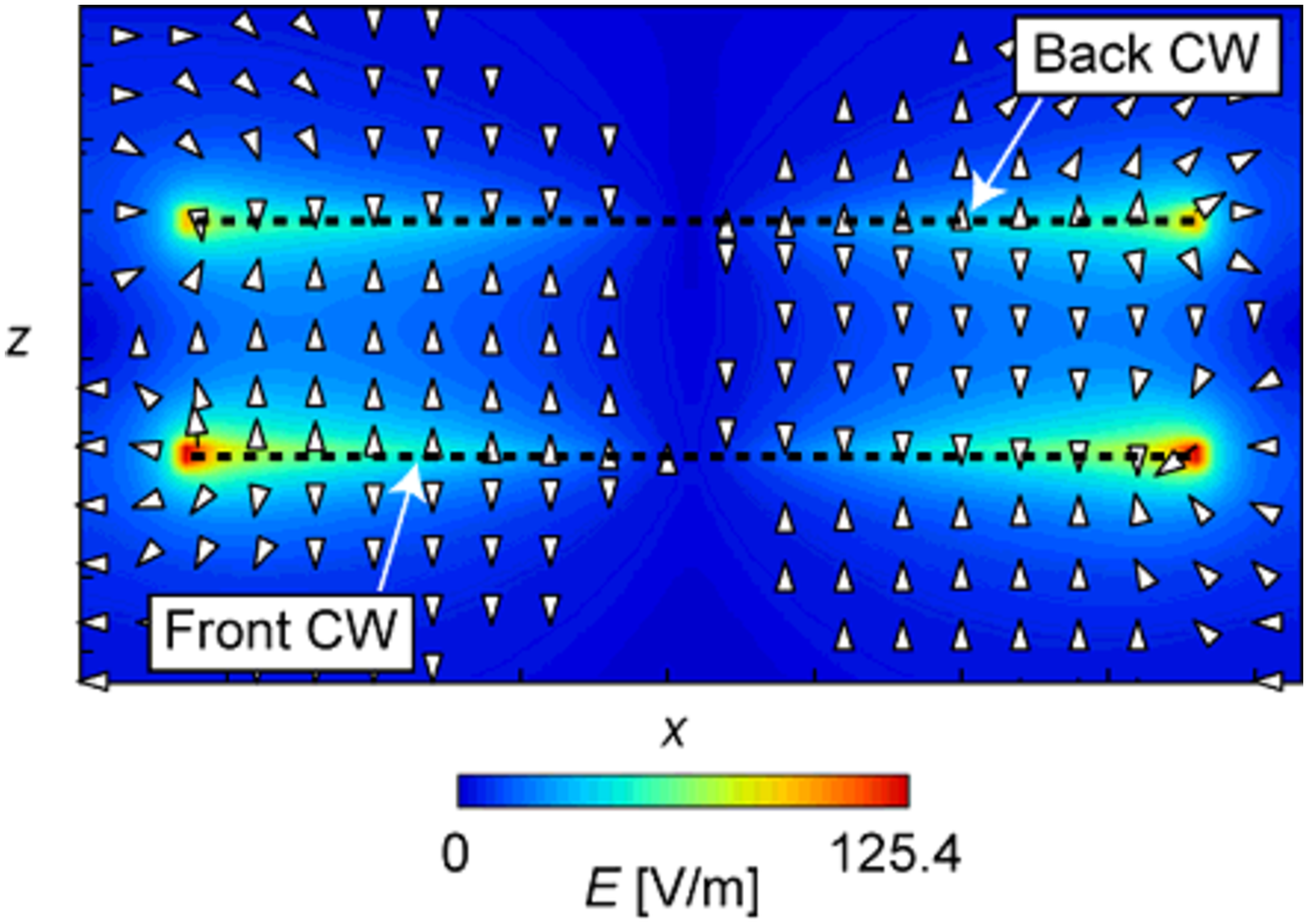}
\end{minipage}
\begin{minipage}{0.5\hsize}
\centering
\includegraphics[width=\linewidth]{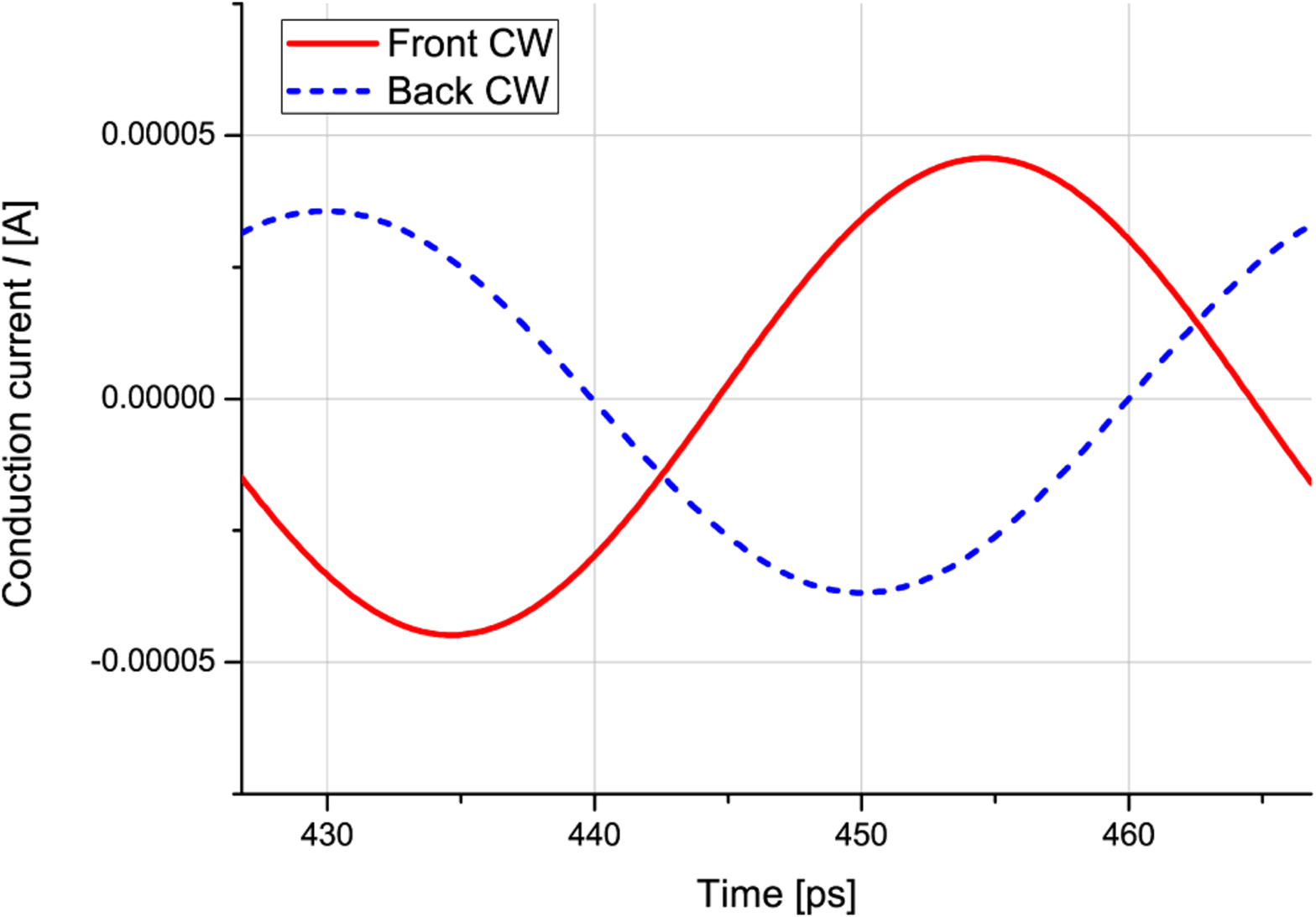}
\end{minipage}
\begin{minipage}{0.5\hsize}
\centering
\includegraphics[width=\linewidth]{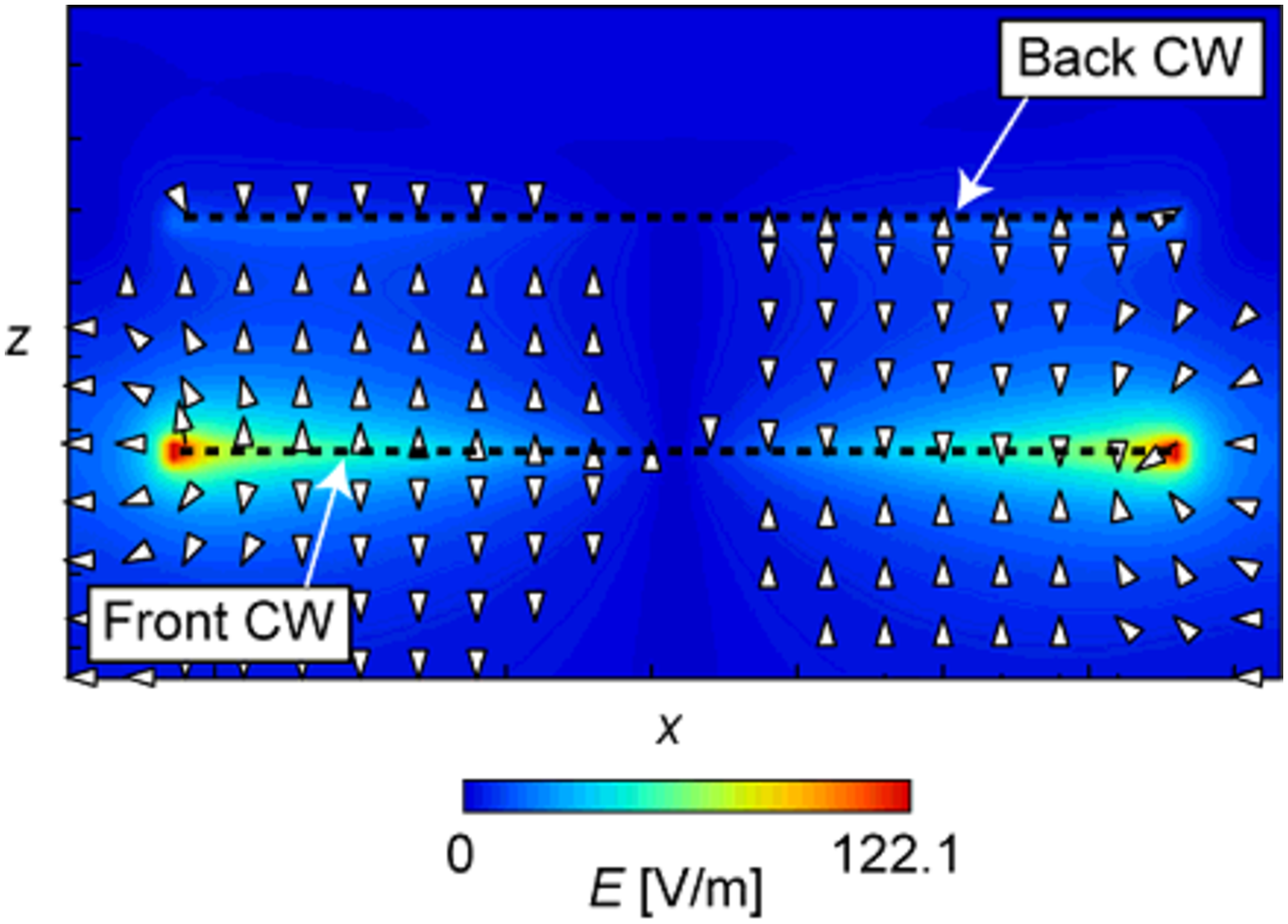}
\end{minipage}
\begin{minipage}{0.5\hsize}
\centering
\includegraphics[width=\linewidth]{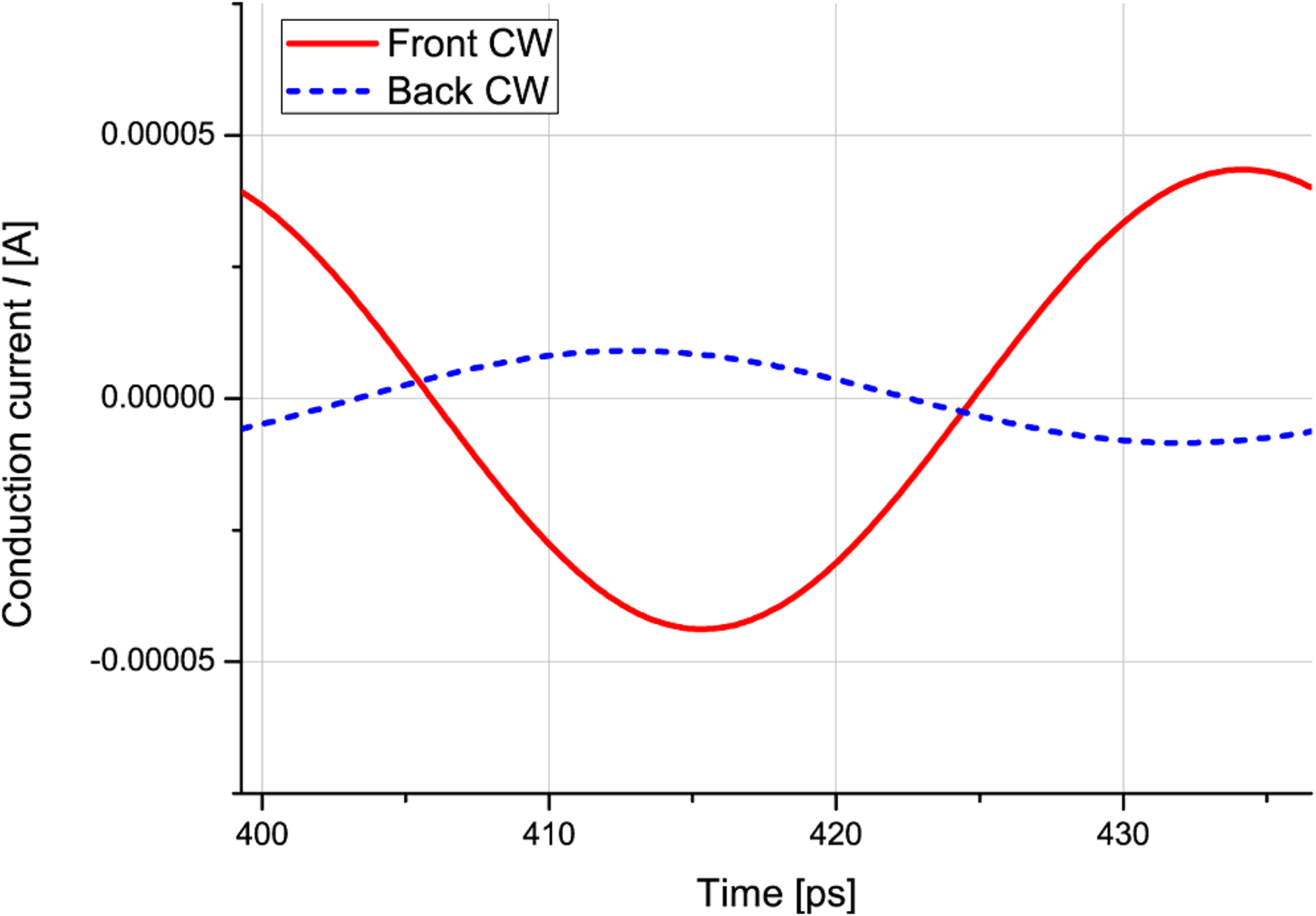}
\end{minipage}
\caption{Default structures of simulated CW metamaterials and their properties. $|S_{11}|$ of (a) lossless single CW and (b) lossless paired CW. (c) Electric field and (d) conduction current of the magnetic resonance frequency (24.89 GHz). (e) Electric field and (f) conduction current of the electric resonance frequency (26.71 GHz). }
\label{fig:modelField}
\end{figure}

The default structures of the simulated CW metamaterials are illustrated in the insets of Figs. \ref{fig:modelField} (a) and (b) for single and paired CWs and the simulations were performed by using the transmission line modelling (TLM) method \cite{TLMbook}. To simplify the situation and consider the absorption effect due only to the conductive loss of the CW metals, the relative permittivity $\varepsilon _r$ of the substrate was set $\varepsilon _r=1$ and all other loss mechanisms were removed. The CW structure had a width of 0.3 mm along the $y$ axis ($H$--field) and a length of 5.1 mm along the $x$ axis ($E$--field). The dimension of the unit cell was $A_x=A_y=6.3$ mm. This structure was modelled using cubical TLM unit cells having edge lengths of $\Delta l=0.075$ mm. The sheet resistance $R$ ($\Omega \Box ^{-1}$) used to quantify the conductive loss was allocated so that the scattering coefficients from the metal are determined by 
\begin{eqnarray}
\Gamma = \frac{R_0-Z_0}{R_0+Z_0} \mathrm{~~~and~~~} T=\frac{2R_0}{R_0+Z_0},
\label{eq:loss}
\end{eqnarray}
where $\Gamma$ and $T$ respectively represent reflection coefficient and transmission coefficient both multiplied by the incident voltage (the incident wave) in each TLM unit cell, $Z_{0}$ is the impedance in vacuum and $R_{0}=RZ_{0}/(R+Z_{0})$. The scattering parameters were calculated using observation planes 15 mm away from the absorber surface and a Gaussian pulse was excited on a plane one cell above the observation plane for the reflection coefficient. Since periodic boundaries were applied for the $xz$ and $yz$ plane boundaries, the CW metamaterial absorber unit modelled was assumed to belong to infinite array on the bottom $xy$ plane. 

The simulated paired CW metamaterials are generally known to exhibit two types of resonances: electric resonances and magnetic resonances \cite{CWeq}. These resonances are explained in Fig. \ref{fig:modelField} where the reflection coefficient, the electric fields and the conduction currents of the CW metamaterials all calculated by TLM method are illustrated. When an incident wave is excited with the electric field parallel to the direction of the CW metals, the conduction currents flow in opposite directions in each CW at the magnetic resonance (see Fig. \ref{fig:modelField} (d)). This is the minimum $|S_{11}|$ at 24.89 GHz in Fig. \ref{fig:modelField} (b). In this case the electric field is anti--parallel at the two edges of the CW metamaterial, Fig. \ref{fig:modelField} (c), and a magnetic field concentrated within the area bounded by the paired CW is generated. This magnetic field plays the role of an artificial magnetic dipole and can lead to a negative value of permeability. On the other hand, at the electric resonance found at around an $|S_{11}|$ peak (26.71 GHz in Fig. \ref{fig:modelField} (b)), conduction currents in the CWs become dominant, resulting in a strong unidirectional electric field at the edges of the CWs, Figs. \ref{fig:modelField} (e) and (f). Similarly with the magnetic resonance this electric field behaves as an artificial electric dipole so as to produce a negative value of permittivity. A similar electric resonance is found in the $|S_{11}|$ peak of the single CW metamaterial (26.65 GHz of Fig. \ref{fig:modelField} (a)) as well. The positions of the two types of the resonances can be manipulated \cite{CWeq,french} and in subsection \ref{ssec:pairedCW} the influence of the resonant frequency positions on the absorptance peak will be discussed. 

\section{Absorptances of single CW metamaterials and paired CW metamaterials}
\label{sec:singleAndPaired}
\subsection{Absorptances of single CW metamaterials}
\begin{figure}[htbp]
\begin{minipage}{0.47\hsize}
\centering
\includegraphics[width=\linewidth]{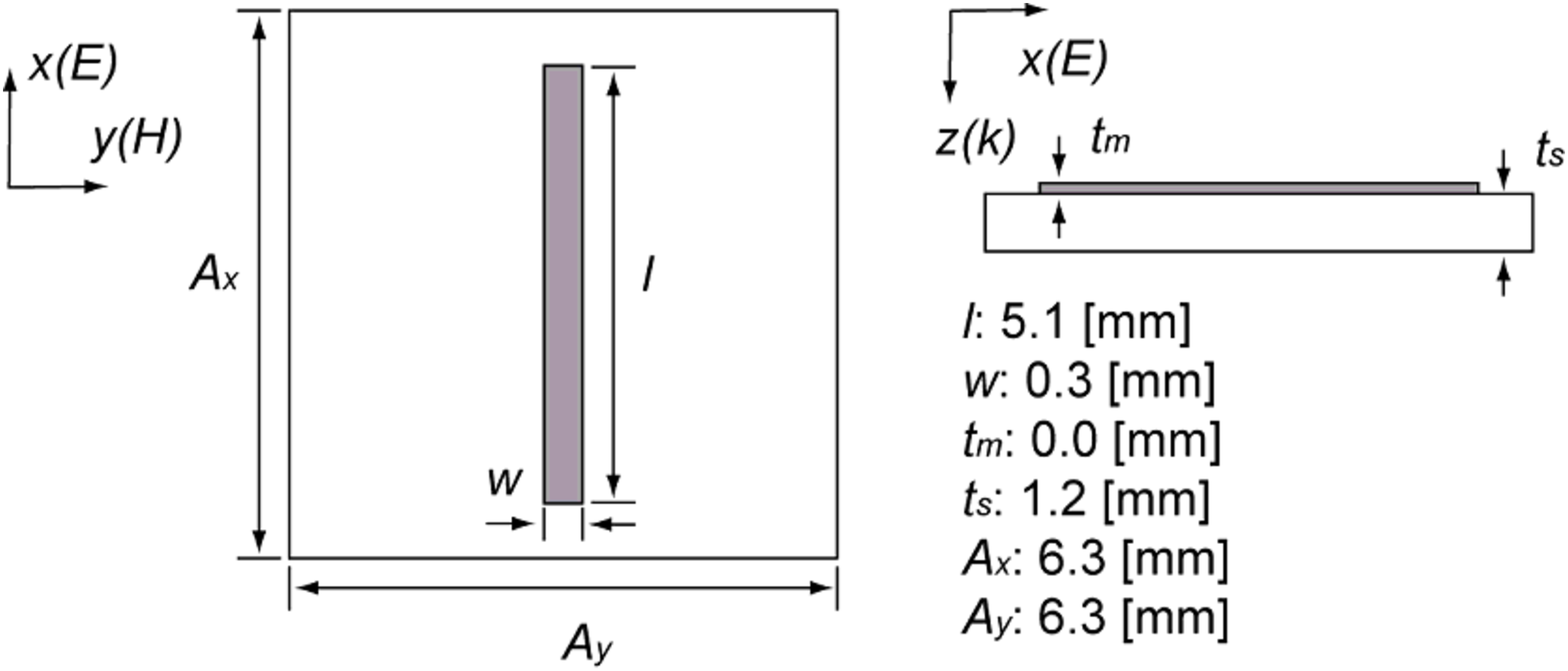}
\end{minipage}
\begin{minipage}{0.53\hsize}
\centering
\includegraphics[width=\linewidth]{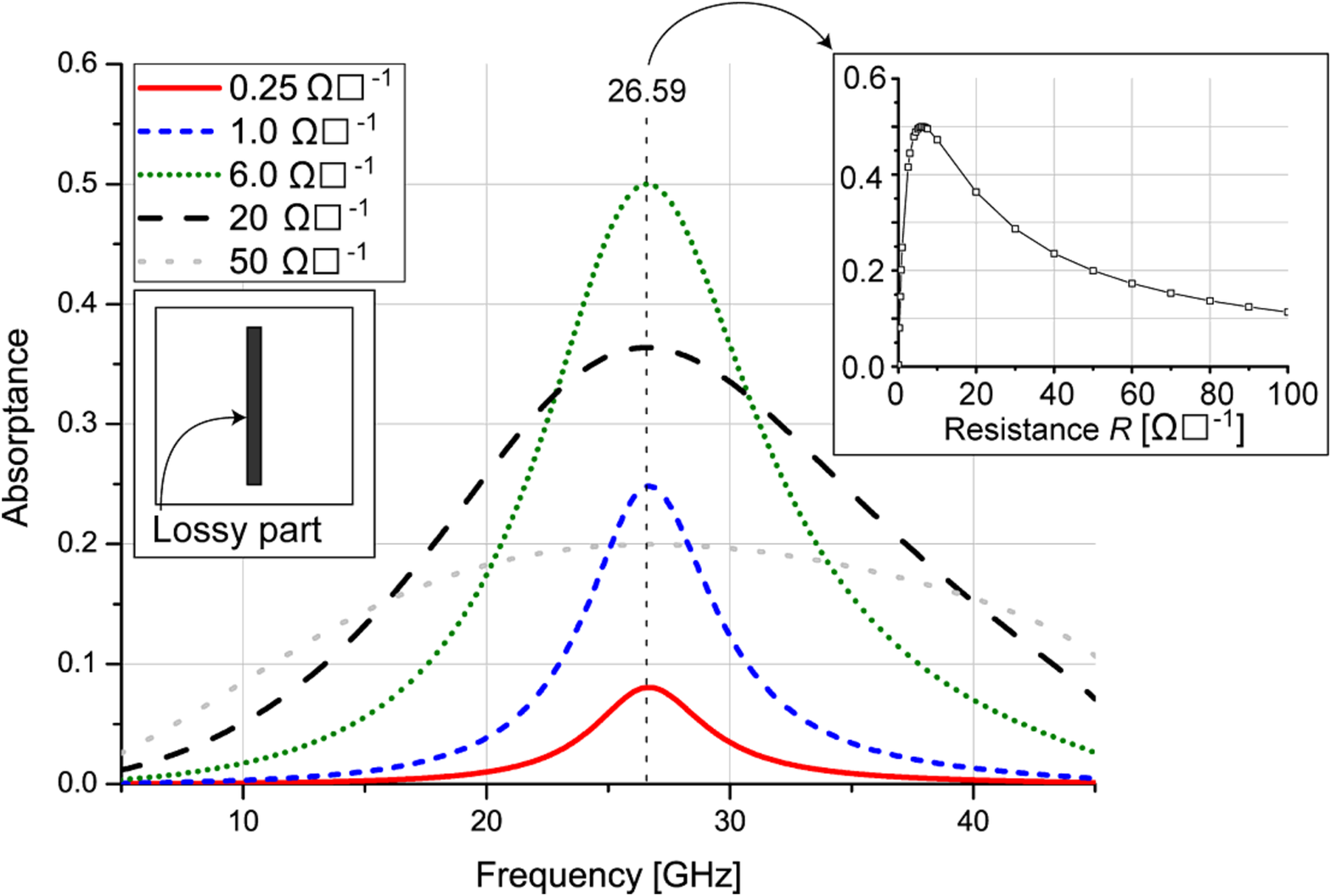}
\end{minipage}
\begin{minipage}{0.5\hsize}
\centering
\includegraphics[width=\linewidth]{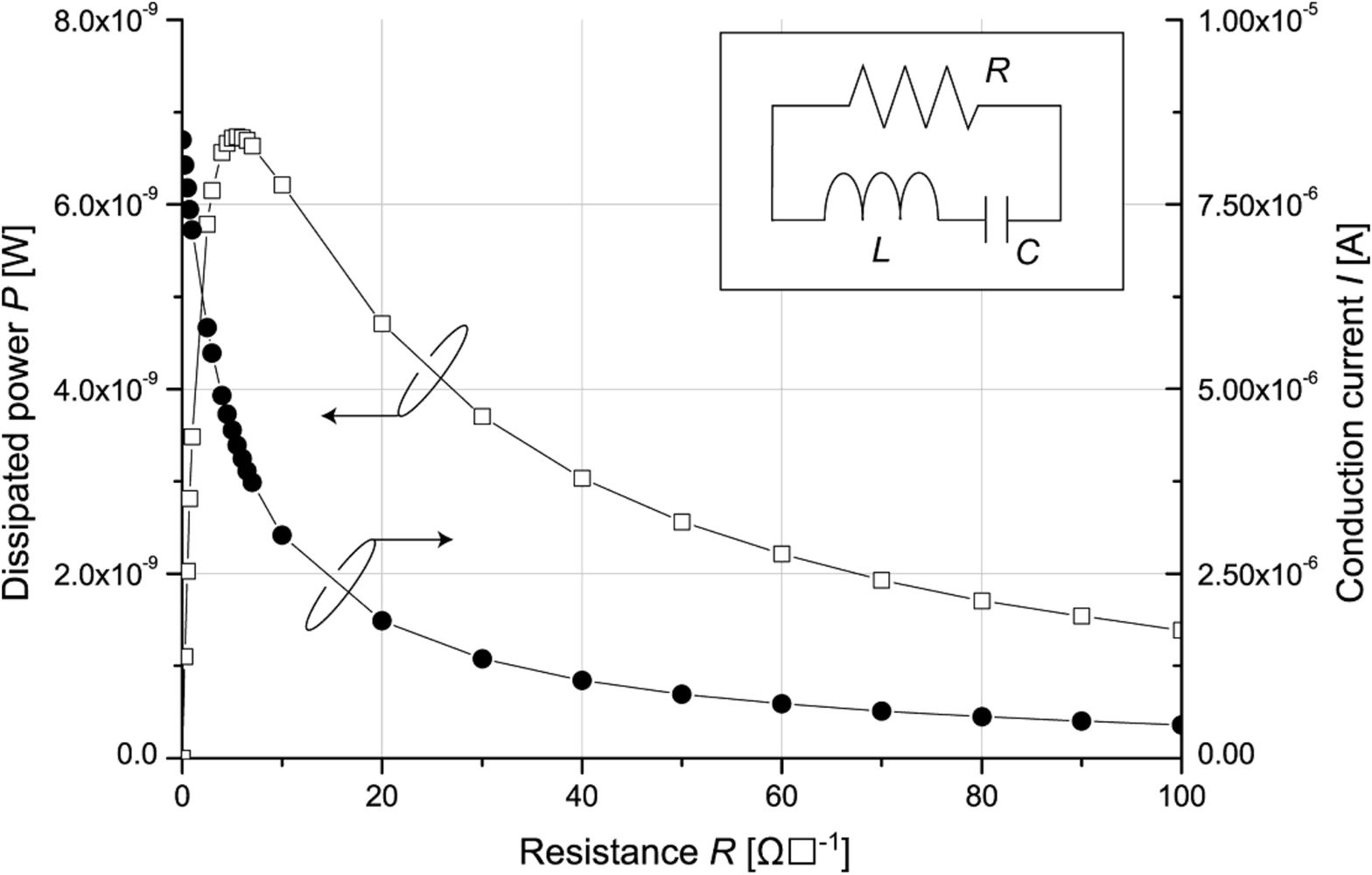}
\end{minipage}
\begin{minipage}{0.5\hsize}
\centering
\includegraphics[width=\linewidth]{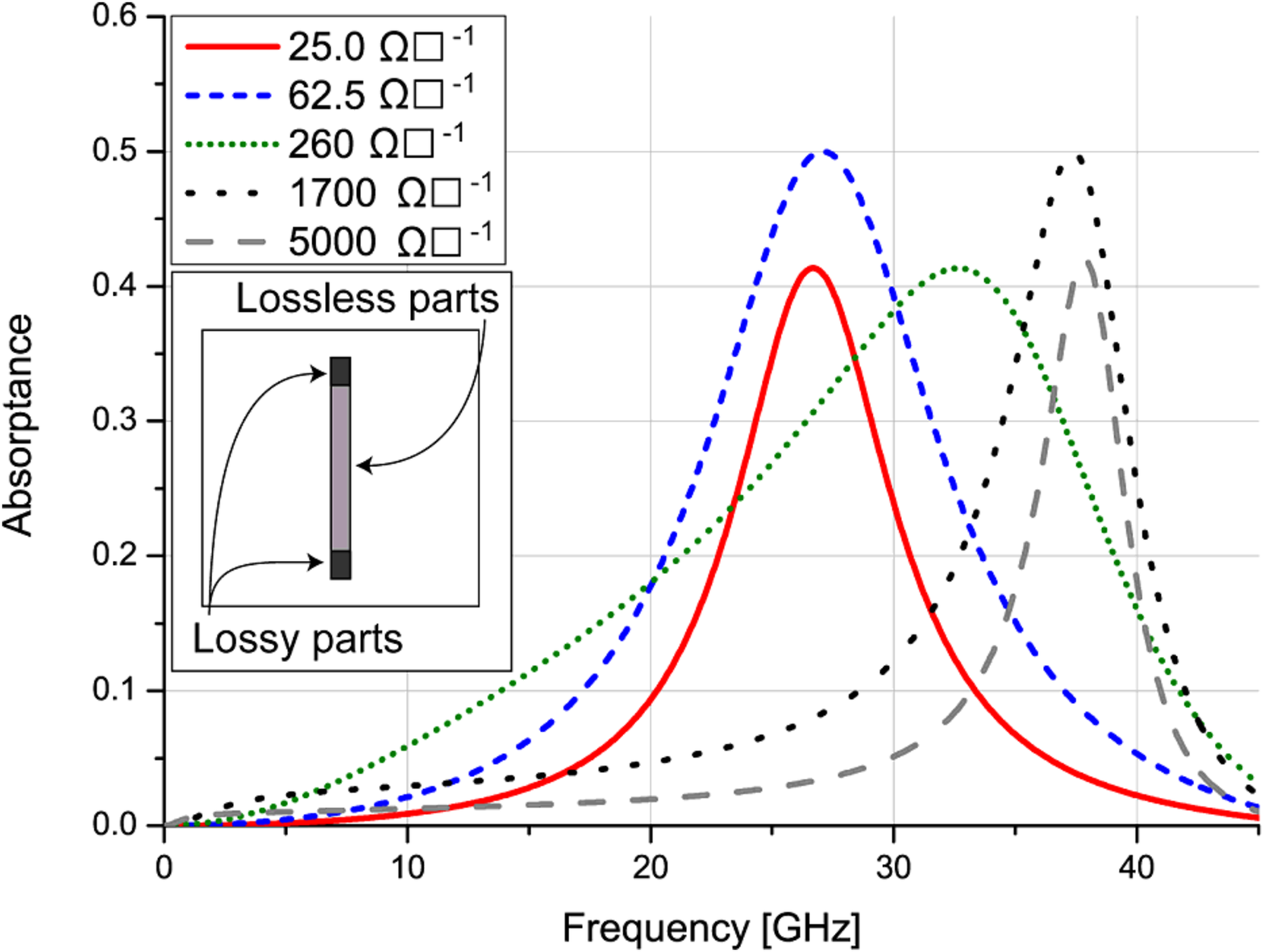}
\end{minipage}
\begin{minipage}{0.5\hsize}
\centering
\includegraphics[width=\linewidth]{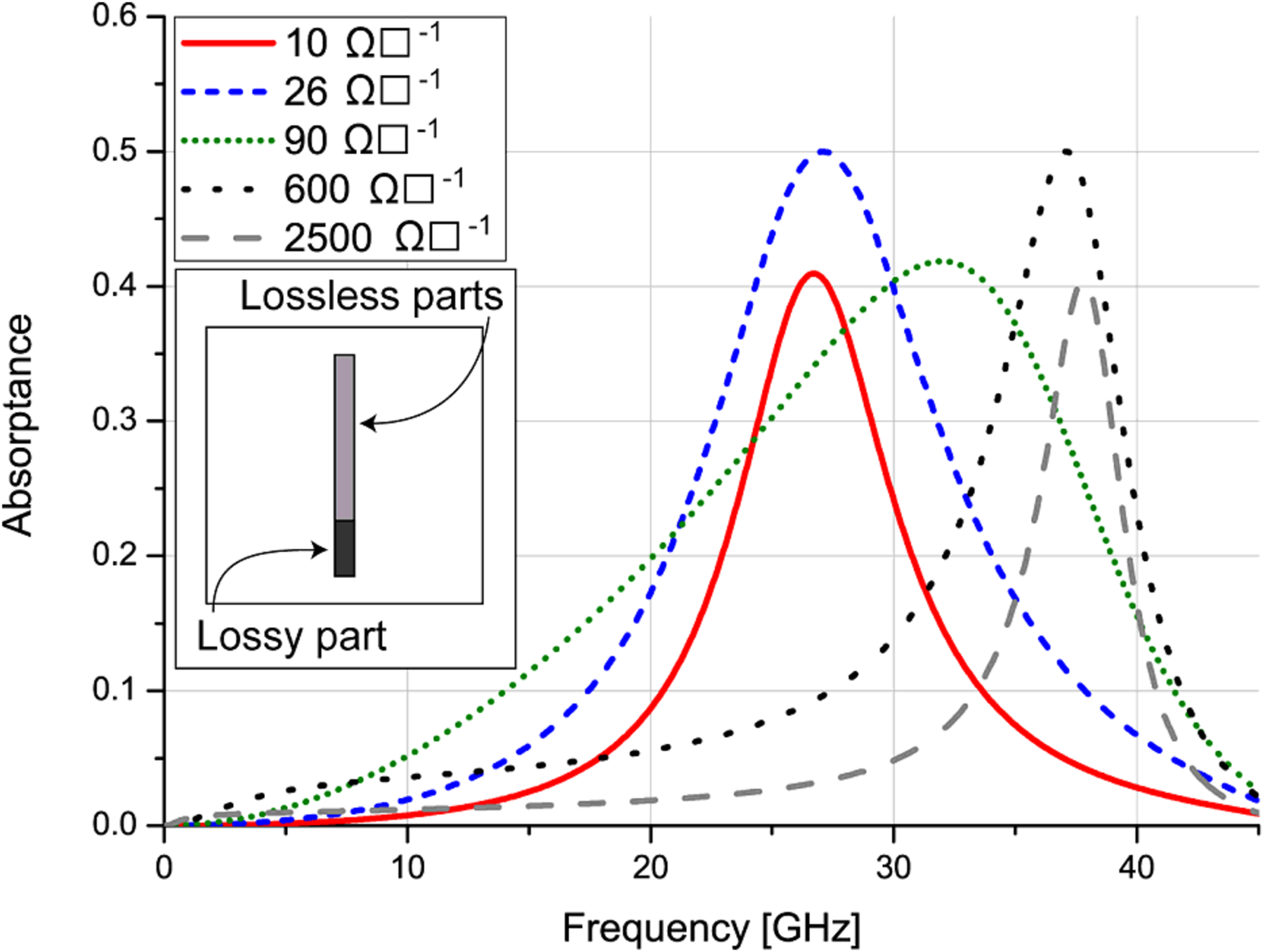}
\end{minipage}
\begin{minipage}{0.5\hsize}
\centering
\includegraphics[width=\linewidth]{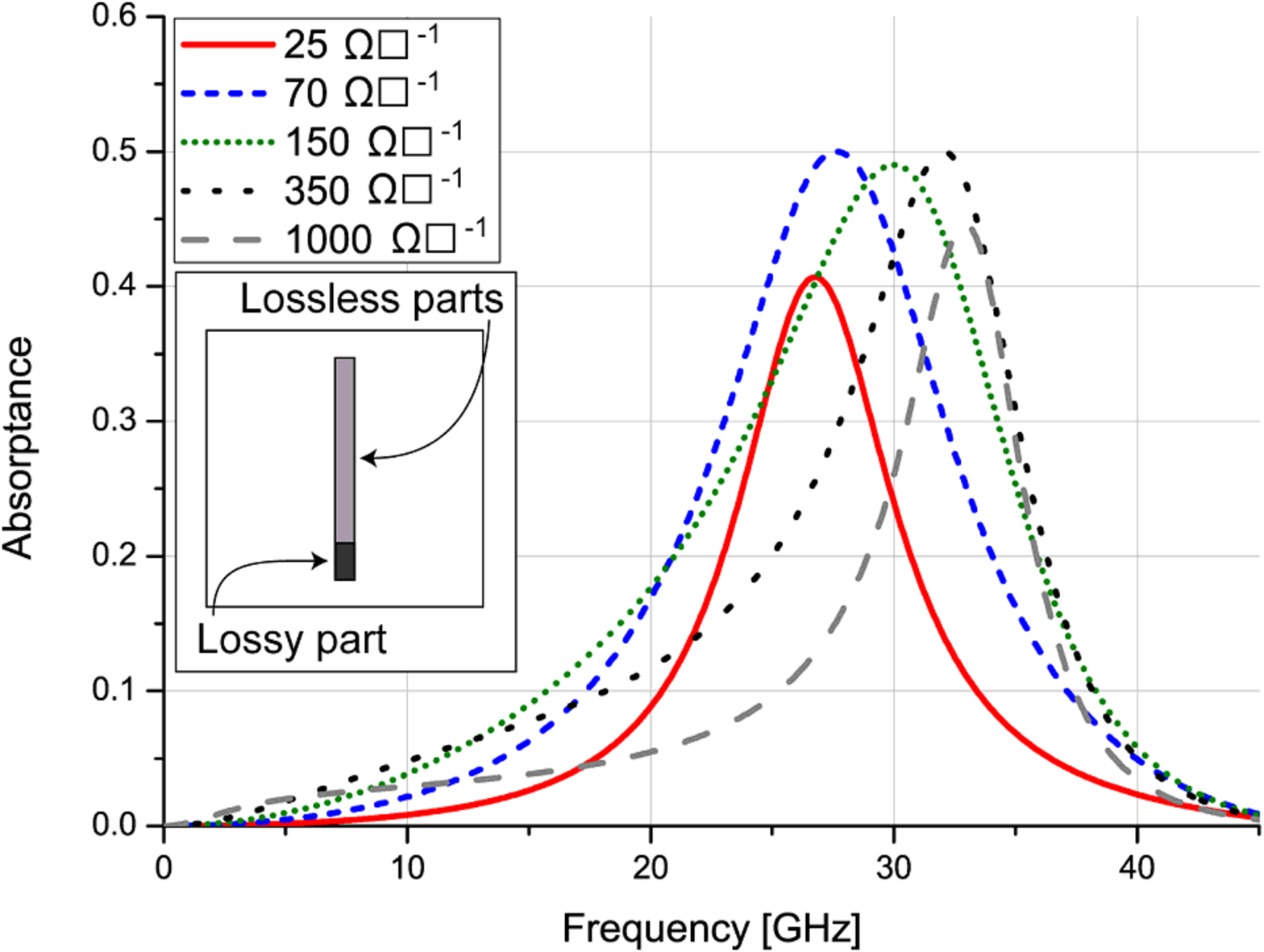}
\end{minipage}
\caption{Absorptances of single CW having constant resistance values. (a) Details of the simulated single CW metamaterial. (b) Absorptance of the single CW composed of only lossy metal. In the inset the absorptance dependence on $R$ at 26.59 GHz is illustrated. (c) The conduction current $I$ and the dissipated power $P$ ($=IR^2$) in the CW with various values of $R$. (d) -- (f) Absorptances of the single CW composed of both lossless metal part and lossy metal part. The lossy metal parts were restricted to both edges (0.9 mm per each edge and 1.8 mm in total) in (d) and one edge in (e) (1.8 mm) and in (f) (1.2 mm). }
\label{fig:allLossySingle}
\end{figure}

This subsection investigates basic absorptance characteristics of single CW metamaterial absorbers. The default size of the periodic structure is illustrated in Fig.\ \ref{fig:allLossySingle} (a). First of all, conductive loss is distributed in one of three ways: allocations to all the metal part, both of the edges (0.9 mm for each edge and 1.8 mm in total) and one of the edges (1.8 mm or 1.2 mm). Secondly, the absorptances of different CW lengths are investigated and they are combined as one metamaterial unit to achieve broadband absorption. In all the simulations the value of the sheet resistance, $R$, is constant. 

In Fig. \ref{fig:allLossySingle} (b) the absorptance $A$ of the single CW composed of only lossy metal is shown for various sheet resistance values. The absorptance was calculated from $A=1-|S_{11}|^2-|S_{21}|^2$. It is found from Fig. \ref{fig:allLossySingle} (b) that the positions of the maximum absorptance are fixed at around 26.59 GHz, which is close to the resonant frequency found in Fig. \ref{fig:modelField} (a). In addition it turns out that the absorptance peak curve reaches the maximum values with a sheet resistance $R=6.0~\Omega \Box ^{-1}$, the inset of Fig. \ref{fig:allLossySingle} (b). This absorptance peak dependence can be easily explained with the equivalent circuit drawn in the inset of Fig. \ref{fig:allLossySingle} (c). At the resonant frequency the circuit impedance becomes $Z=R$. In terms of the conduction current $I$, numerical simulations revealed its dependence on $R$ as the curve with closed circles shown in Fig. \ref{fig:allLossySingle} (c). Therefore, the power dissipated in the circuit, $P$ ($=I^2R$), as illustrated by the curve with open squares in Fig. \ref{fig:allLossySingle} (c), well corresponds to the result in the inset of Fig. \ref{fig:allLossySingle} (b). 

Similar maximum absorptance peaks were obtained when the conductive loss area was restricted to one or both of the edges of the CW metal (see Figs. \ref{fig:allLossySingle} (d) to (f)). However, further increase of the loss amount exhibited different absorptance peaks at higher frequencies. These absorption positions were expected to correspond to resonances of CWs shortened by the lossy metal part(s), since the conductive losses were too high causing a highly distorted current distribution. Numerical simulations confirmed that the 3.3 mm--length CW, whose length correspond to the lossless parts of Figs. \ref{fig:allLossySingle} (d) and (e), has resonant frequency at 38.06 GHz, which is in the vicinity of the absorptance peaks at the higher frequencies in Figs. \ref{fig:allLossySingle} (d) and (e). Similarly the 3.9 mm--length CW, whose metal length is same with the lossless part of Fig.\ \ref{fig:allLossySingle} (f), was found to have resonant frequency at 33.55 GHz, which is also close to another absorptance peak of Fig.\ \ref{fig:allLossySingle} (f). Although the absorptance effect obtained from these structures is weak, one point noticed from Figs.\ \ref{fig:allLossySingle} (e) and (f) is that the decrease of the absorptance at the frequency range between the two peaks can be reduced by restricting the lossy metal part to the narrow area. 

%

\begin{figure}[htbp]
\begin{minipage}{0.5\hsize}
\centering
\includegraphics[width=\linewidth]{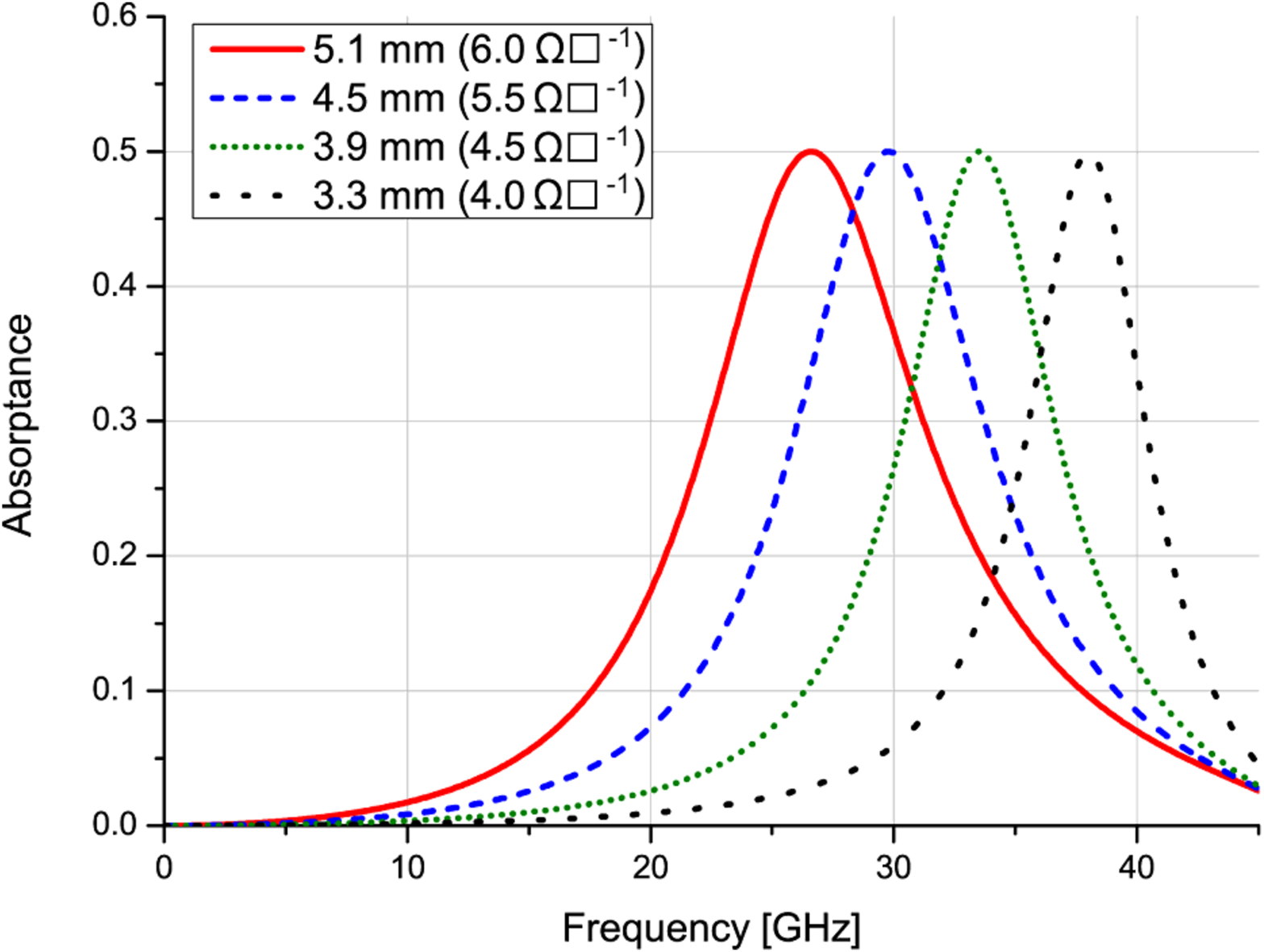}
\end{minipage}
\begin{minipage}{0.5\hsize}
\centering
\includegraphics[width=\linewidth]{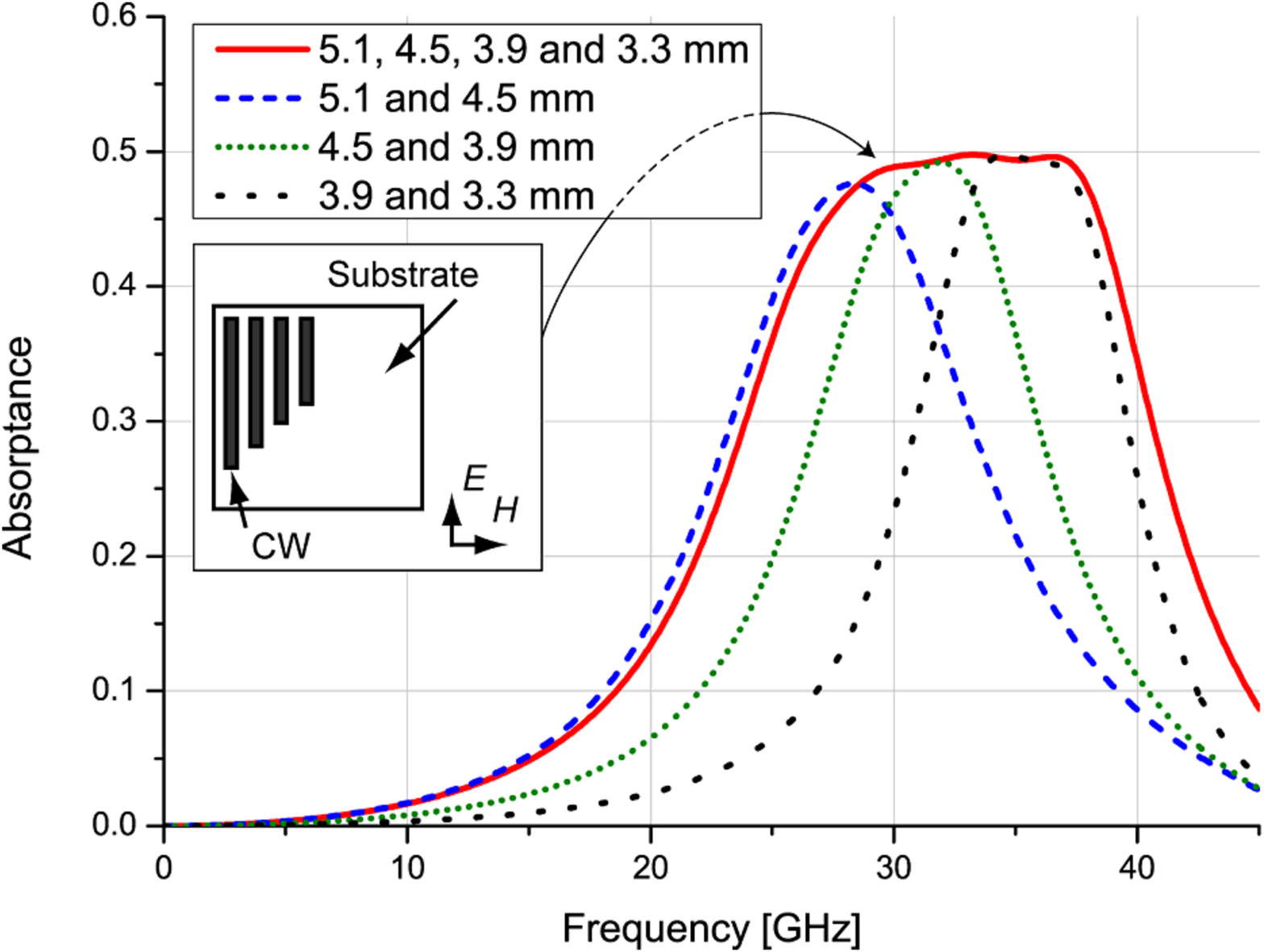}
\end{minipage}
\caption{Absorptance of different lengths of single CWs (a) and absorptance of structure composed of part or all of the CWs (b). The lengths and the resistances in the legends represent the CW length and $R$, respectively. }
\label{fig:diffLength}
\end{figure}

Secondly, the absorptances of different CW lengths were calculated. The calculation results are summarised in Fig. \ref{fig:diffLength} (a) where the simulated CWs had lengths of 3.3 to 5.1 mm by 0.6 mm steps. It is found from Fig. \ref{fig:diffLength} (a) that the absorptance peak depends on the CW length. Next several of the CWs used in Fig.\ \ref{fig:diffLength} (a) were combined as one metamaterial unit and the resultant absorptance curves are illustrated in Fig. \ref{fig:diffLength} (b). The inset in Fig. \ref{fig:diffLength} (b) describes a structure consisting of four CWs. In this structure the distance between the CWs was 0.3 mm and one end of the CWs were positioned at a same $x$ axis position (0.3 mm from one of $xz$ plane boundaries). In the other structures only those lengths noted were included. Compared to the results in Fig. \ref{fig:diffLength} (a), the results in Fig. \ref{fig:diffLength} (b) illustrate broadband behaviour. 

\subsection{Absorptances of paired CW metamaterials}
\label{ssec:pairedCW}
\begin{figure}[htbp]
\begin{minipage}{0.48\hsize}
\centering
\includegraphics[width=\linewidth]{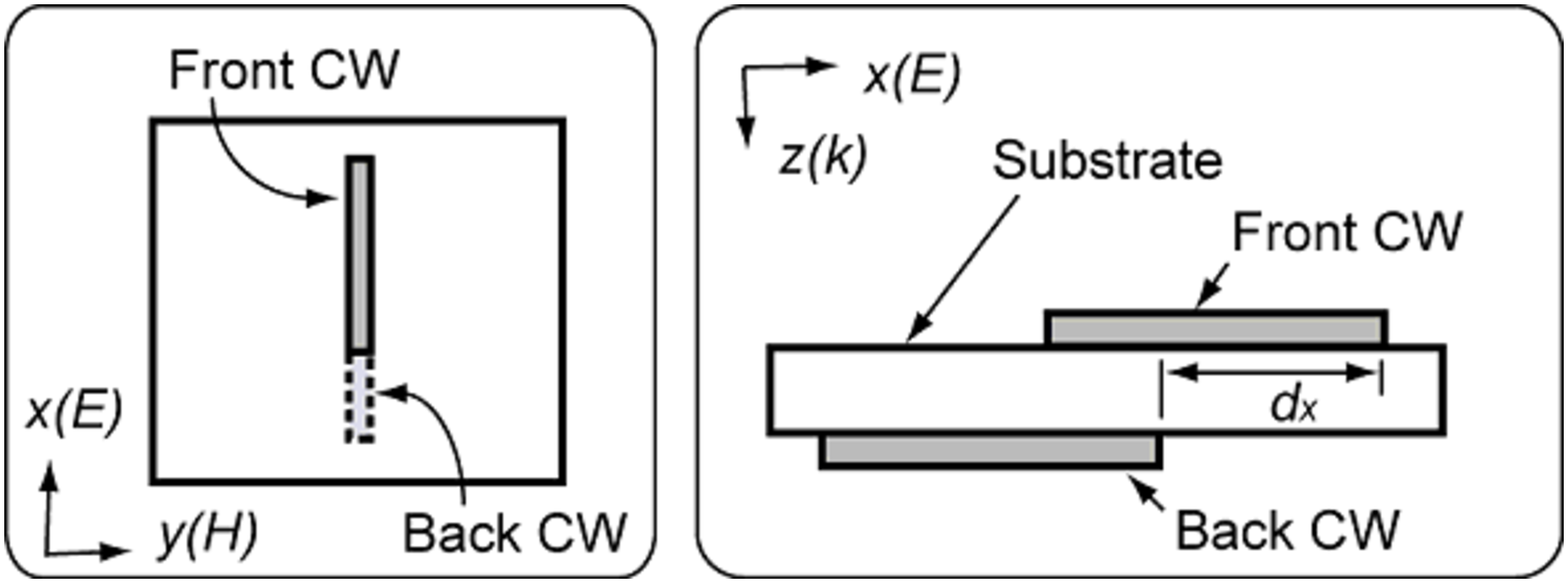}
\end{minipage}
\begin{minipage}{0.48\hsize}
\centering
\includegraphics[width=\linewidth]{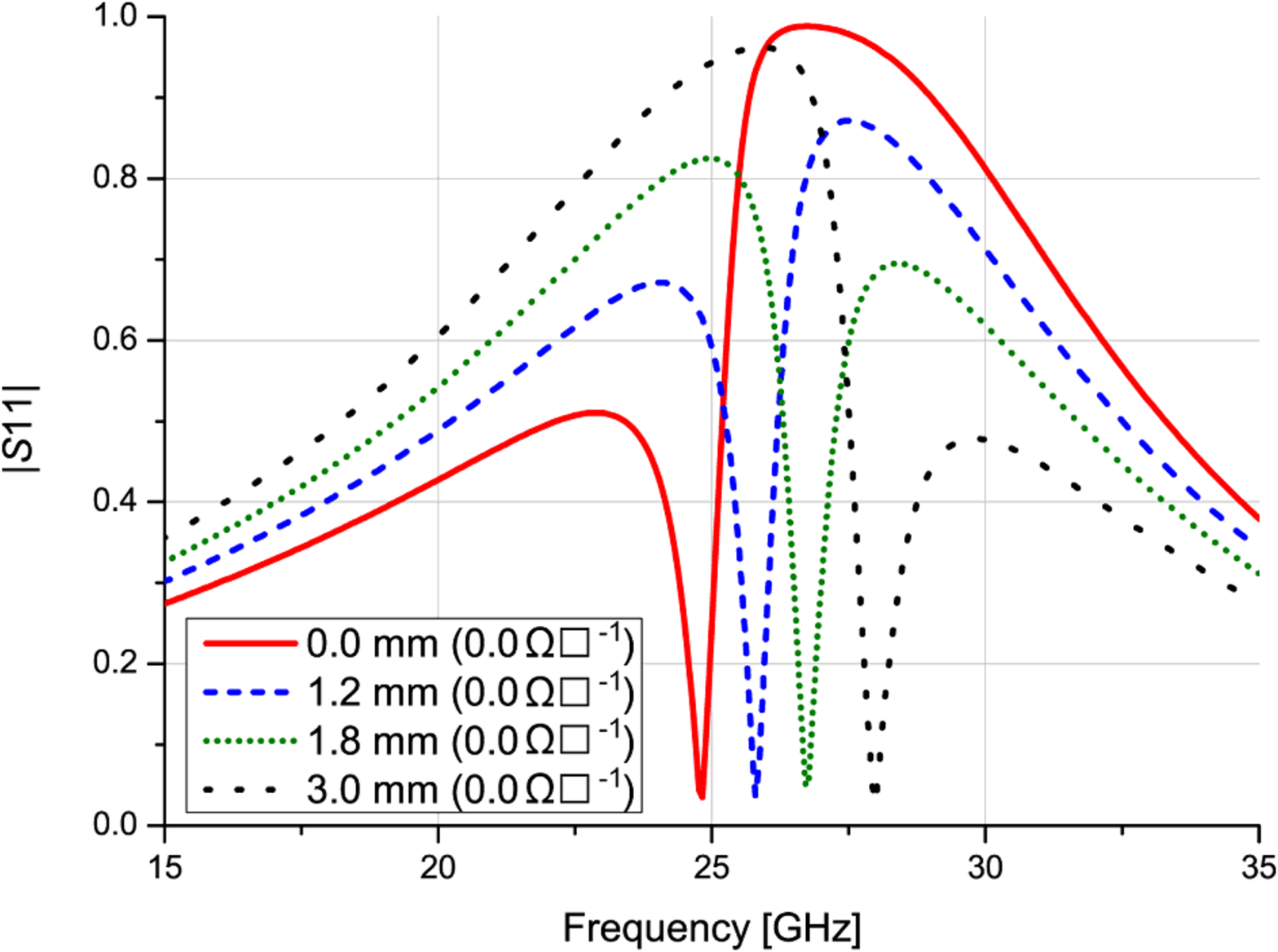}
\end{minipage}
\begin{minipage}{0.48\hsize}
\centering
\includegraphics[width=\linewidth]{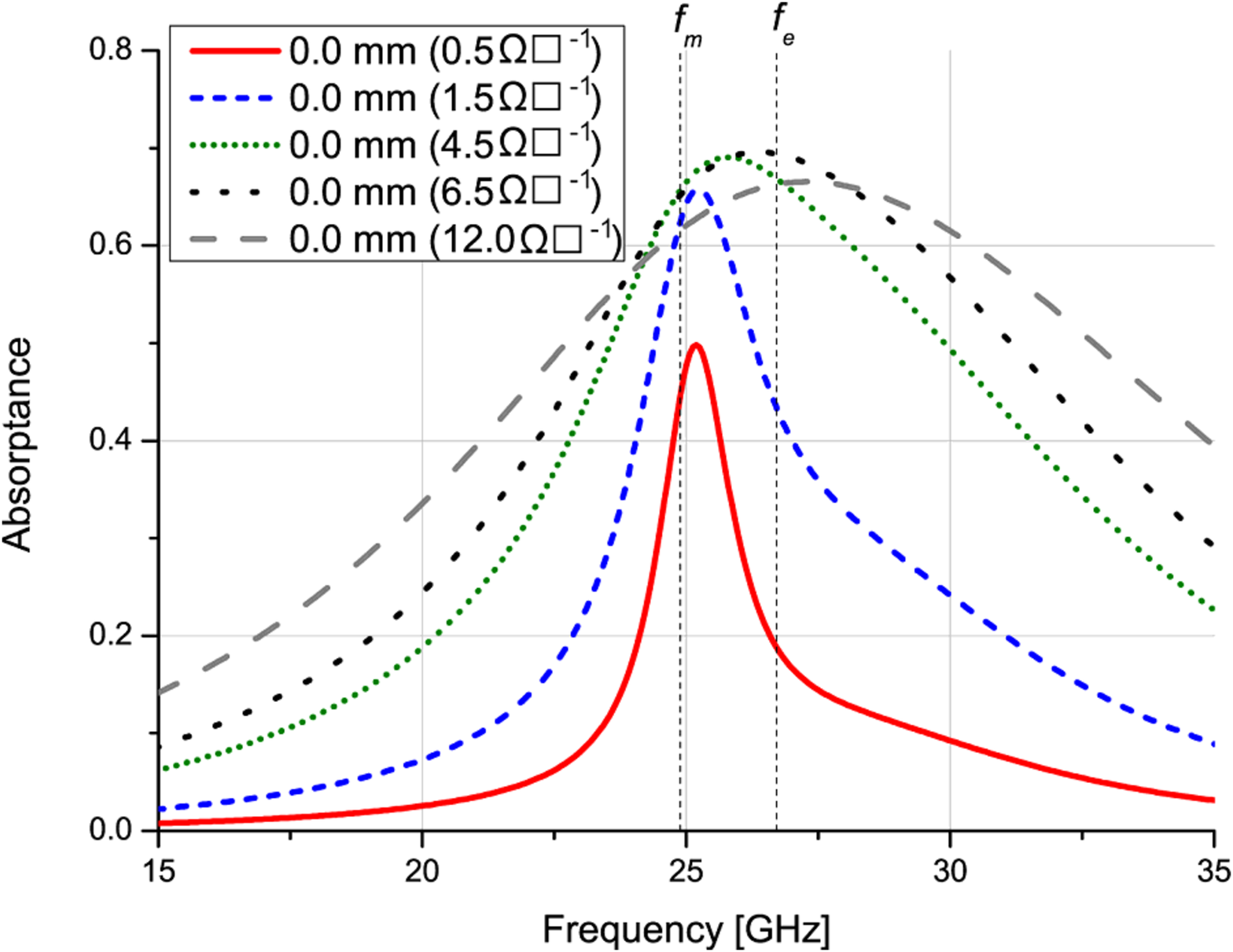}
\end{minipage}
\begin{minipage}{0.48\hsize}
\centering
\includegraphics[width=\linewidth]{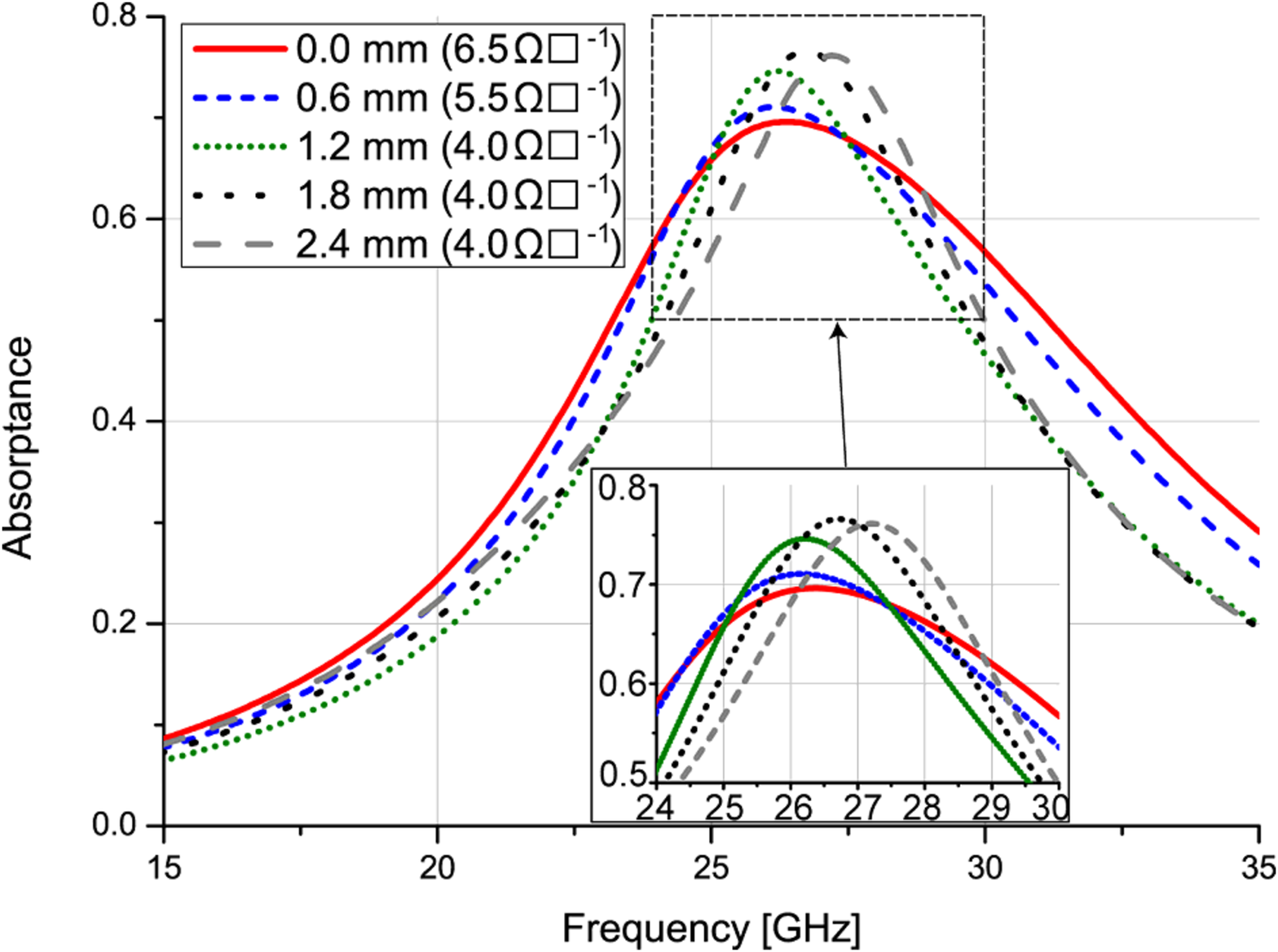}
\end{minipage}
\begin{minipage}{0.32\hsize}
\centering
\includegraphics[width=\linewidth]{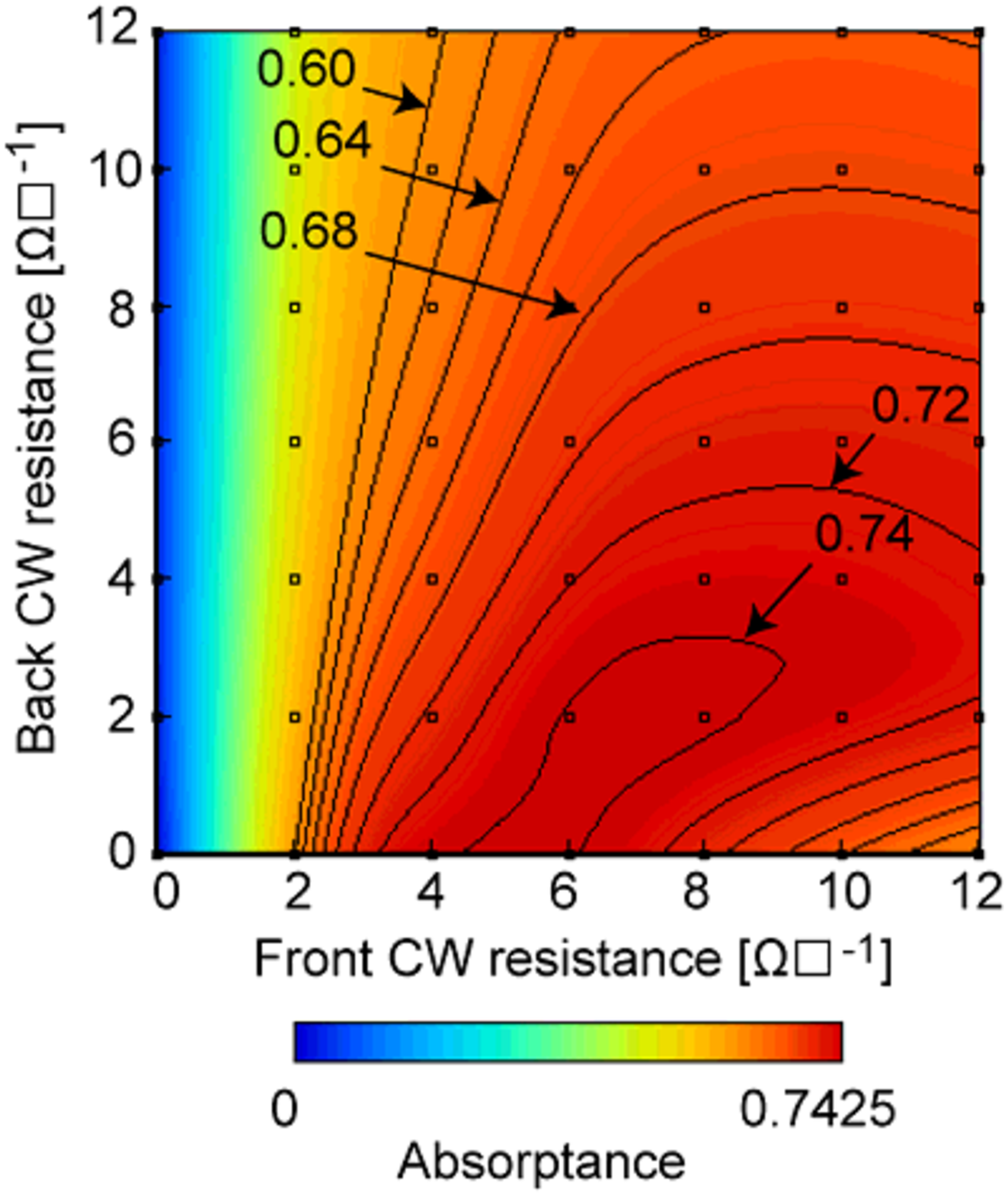}
\end{minipage}
\begin{minipage}{0.32\hsize}
\centering
\includegraphics[width=\linewidth]{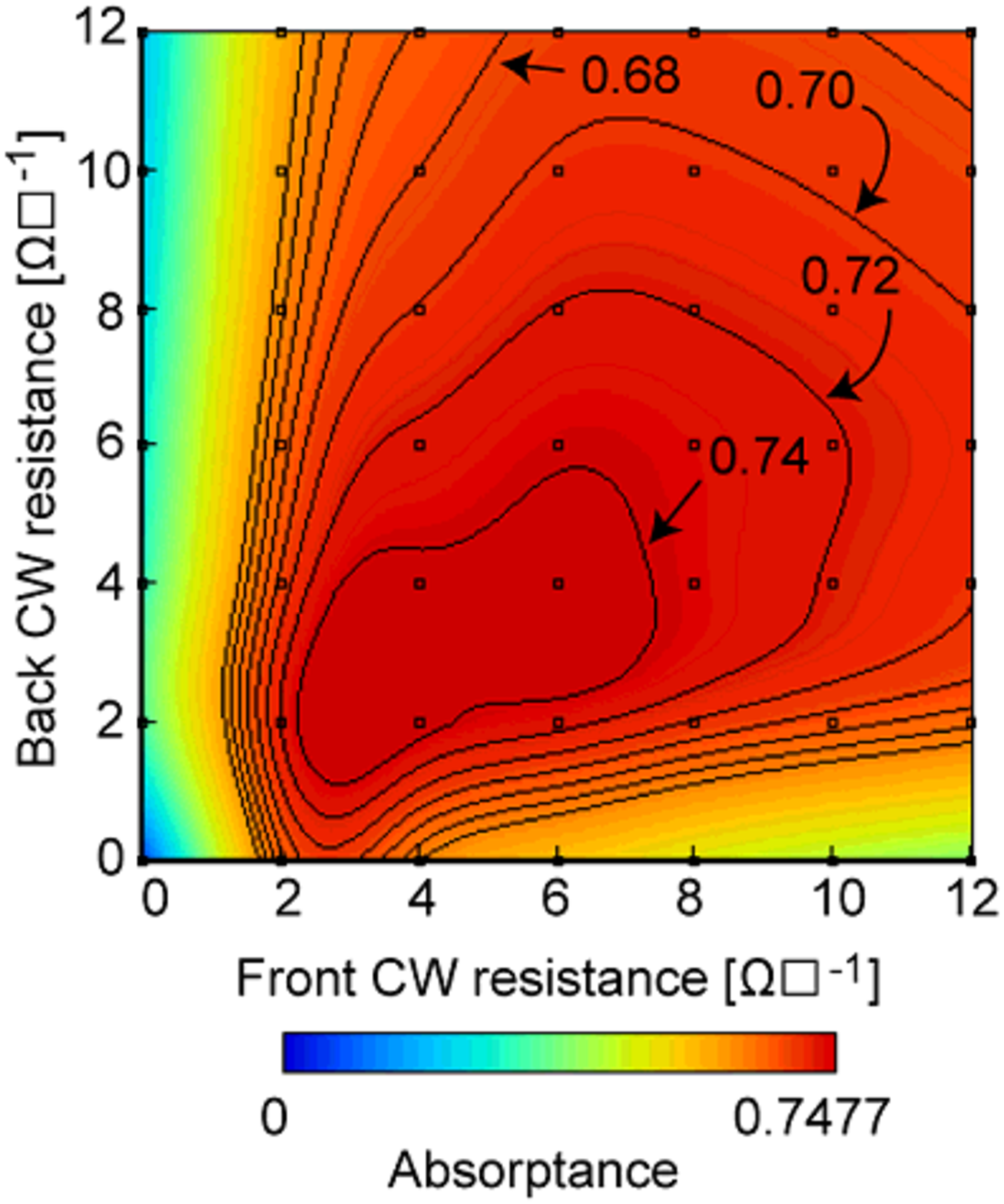}
\end{minipage}
\begin{minipage}{0.32\hsize}
\centering
\includegraphics[width=\linewidth]{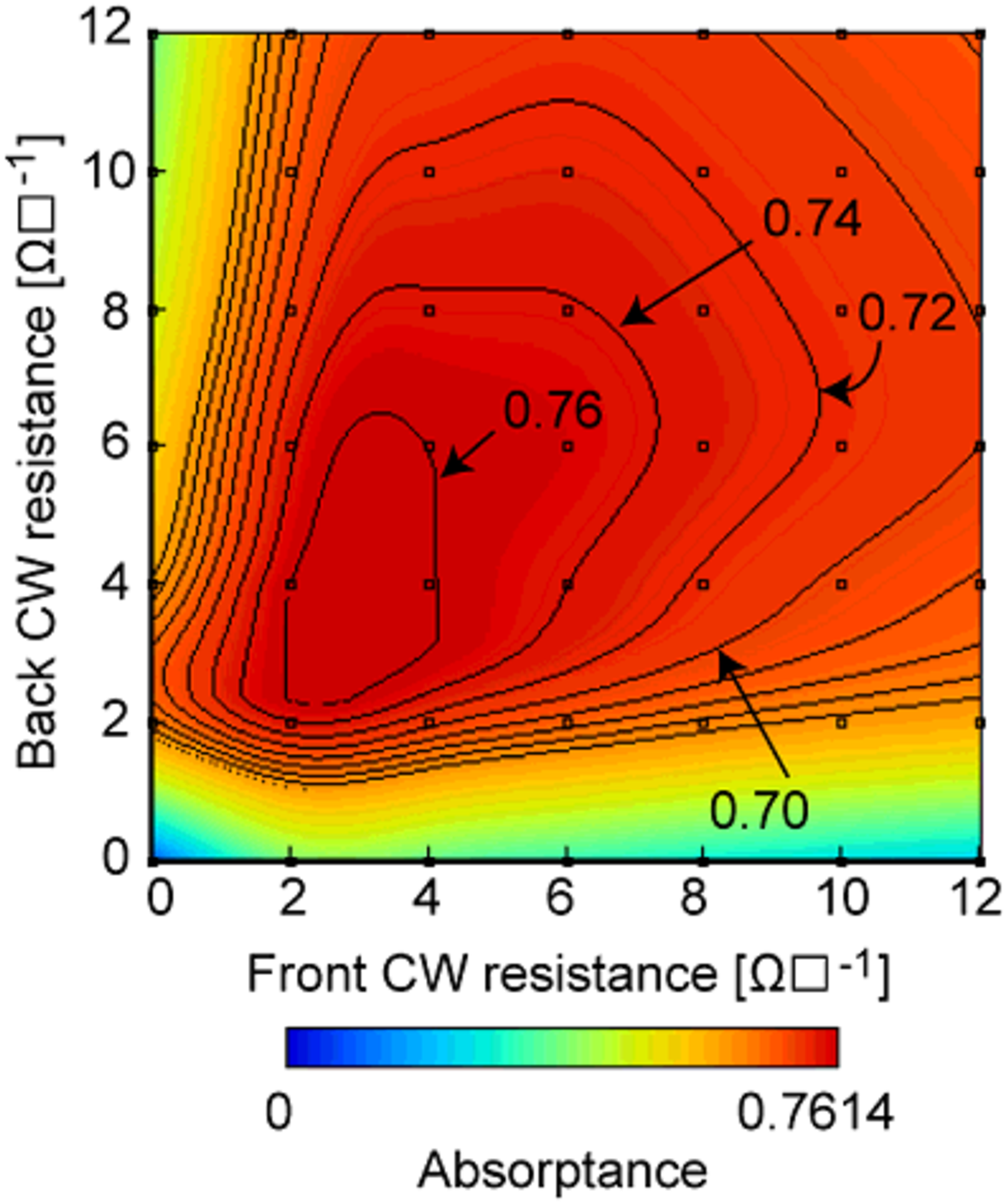}
\end{minipage}
\caption{Characteristics of paired CW. Geometrical asymmetry was introduced as is described in (a). (b) $|S_{11}|$ of the lossless paired CW having the various geometrical offset length $d_x$. (c) Absorptance of the symmetrically paired CW with various $R$. (d) Absorptances of the paired CW with various $d_x$. From (b) to (d) $d_x$ is expressed in each legend. In these figures same values of $R$ are used for the front CW and the back CW. In (e) to (g) various combinations of $R$ are applied for both CWs. (e), (f) and (g) show the absorptances of no offset at 26.35 GHz, 1.2 mm offset at 26.23 GHz and 2.4 mm offset at 27.20 GHz, respectively. The dots in the figures represent the calculated sheet resistance patterns and the intermediate values were estimated by using spline interpolation. }
\label{fig:offset}
\end{figure}

This subsection examines absorptance characteristics of paired CW metamaterials. As was explained in section \ref{sec:material}, the paired CW metamaterials have two types of resonance: electric resonance and magnetic resonance. These resonant frequency positions can be manipulated and in \cite{french}, for example, one of the CW pairs is shifted a distance $d_x$ (see Fig.\ \ref{fig:offset} (a)). In this case the magnetic resonance frequency is increased, while the electric resonance frequency is reduced. This effect is shown in Fig.\ \ref{fig:offset} (b) where the dependence of the reflection coefficient curves on the various offset lengths $d_x$ are shown. It is known that, when the magnetic resonance frequency is greater than the electric resonance frequency, a NRI is obtained in the paired CW metamaterial due to overlap of the negative permeability with the negative permittivity \cite{CWeq}. This paper utilises the geometrical asymmetry introduced in \cite{french} to change the resonance positions and implement loss in the structure so that the absorption characteristics in the paired CW metamaterial absorber are investigated in more detail. 

Before the asymmetrically paired CW is simulated, the absorption characteristics of the symmetrically paired CWs are investigated. The calculation results are summarised in Fig.\ \ref{fig:offset} (c) in which various sheet resistance values, $R$, are applied equally to the front and back CWs. In Fig.\ \ref{fig:offset} (c) we have marked the electric and magnetic resonant frequencies ($f_e$ and $f_m$) of the symmetrically paired CW shown in Fig.\ \ref{fig:offset} (b) and it is seen that the absorptance peak shifts from $f_m$ to $f_e$ as $R$ increases. It is also found that the symmetrically paired CW shows the stronger absorption ($A\simeq 0.696$) than that of the single CWs ($A\simeq 0.500$ in Fig. \ref{fig:allLossySingle}). 

Next, the absorptance of the paired CW metamaterial absorber with various values of $d_x$ was optimised. The calculation results are summarised in Table \ref{tab:offset} and Fig. \ref{fig:offset} (d). Again, the same resistance values were applied for both front CW and back CW. It turns out in this case that the maximum absorptance of the symmetrically paired CW is increased by nearly 10 \% for $d_x = 1.8$ mm. 

Moreover the absorptance of the paired CW was further enhanced by using different sheet resistance values for the front CW and the back CW. In Figs.\ \ref{fig:offset} (e), (f) and (g) the absorptances for $d_x=$ 0.0 mm, 1.2 mm and 2.4 mm are illustrated. Each distribution was calculated at the frequency corresponding to each absorptance peak; 26.35, 26.23 and 27.20 GHz, respectively. In these figures, the sheet resistance values were varied from 0 to 12 $\Omega \Box ^{-1}$ in 2 $\Omega \Box ^{-1}$ steps. The intermediate values between the calculated results were estimated using spline interpolation. Despite the coarse resolution the simulations confirm that the absorptance is further increased in each case by the use of different sheet resistance values. In addition the distribution of the absorptance depends on the offset length $d_x$. Although the absorptance of the symmetrically paired CW metamaterial was enhanced by the manipulation of the resonant frequency positions and by use of different resistance values, the maximum value was still far from perfect absorption (i.e. $A\sim 1.0$). In the next section the absorptance of the CW metamaterial absorber is significantly improved and approaches perfect absorption by placing a single CW on a perfect electric conductor (PEC) wall. 

\begin{table}[htbp]
\caption{Absorptance peaks of paired CW with various $x$ axis offset $d_x$.}
\label{tab:offset}
\centering
\begin{tabular}{c|ccc}
\hline
$d_x$ [mm]& Absorptance & Frequency [GHz] & $R$ [$\Omega \Box ^{-1}$]\\
\hline
0.0 & 0.696 & 26.35 & 6.5\\
0.6& 0.711 & 26.17 & 5.5 \\
1.2& 0.746 & 26.23 & 4.0  \\
1.8& 0.766 & 26.71 & 4.0 \\
2.4& 0.761 & 27.20 & 4.0 \\
3.0& 0.754 & 27.45 & 4.0 \\
3.6& 0.757 & 27.39 & 4.0 \\
\hline
\end{tabular}
\end{table}

\section{Absorptances of single CW metamaterials placed on PEC}
\label{sec:CWpec}
\begin{figure}[htbp]
\begin{minipage}{0.5\hsize}
\centering
\includegraphics[width=\linewidth]{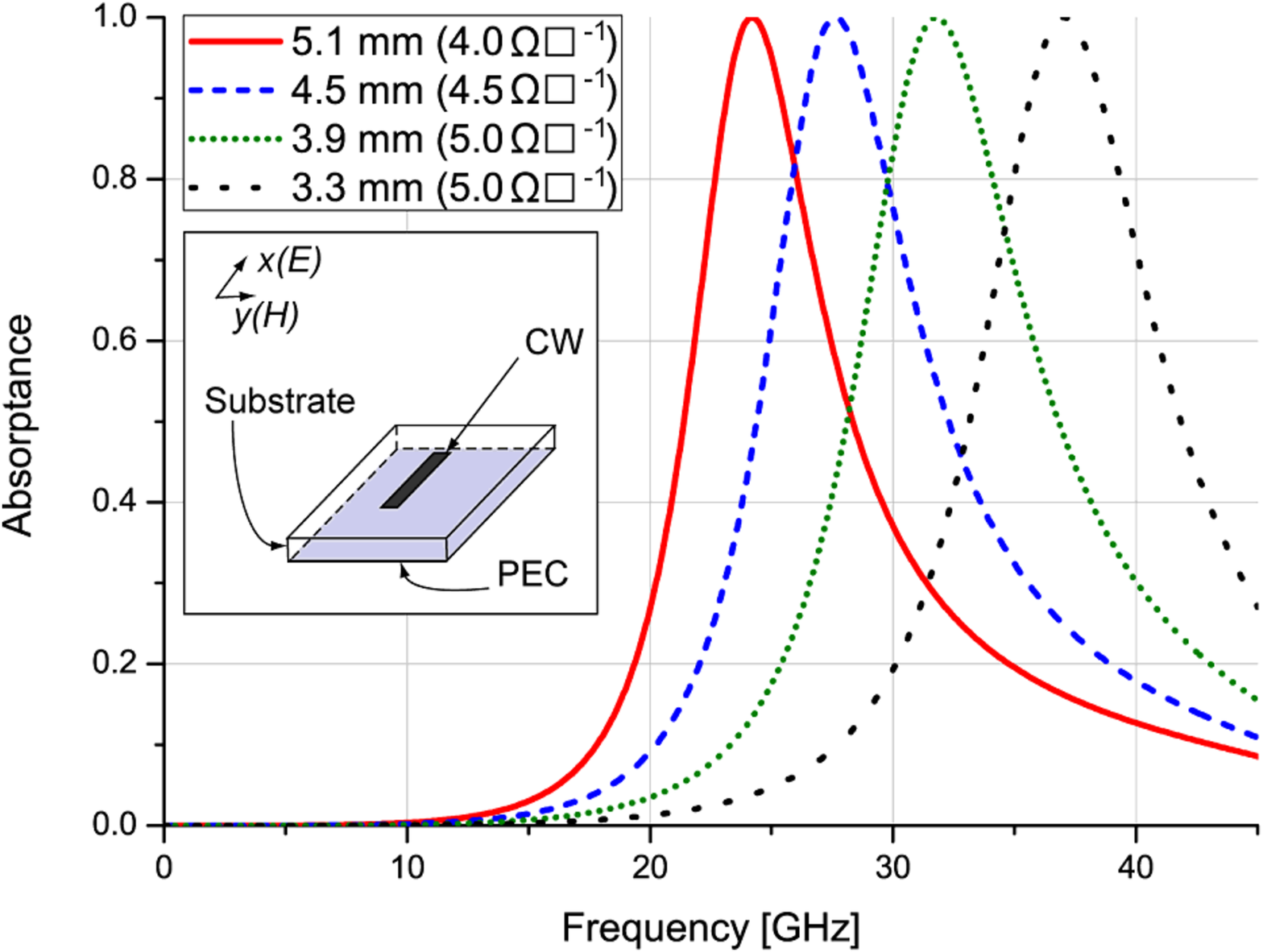}
\end{minipage}
\begin{minipage}{0.5\hsize}
\centering
\includegraphics[width=\linewidth]{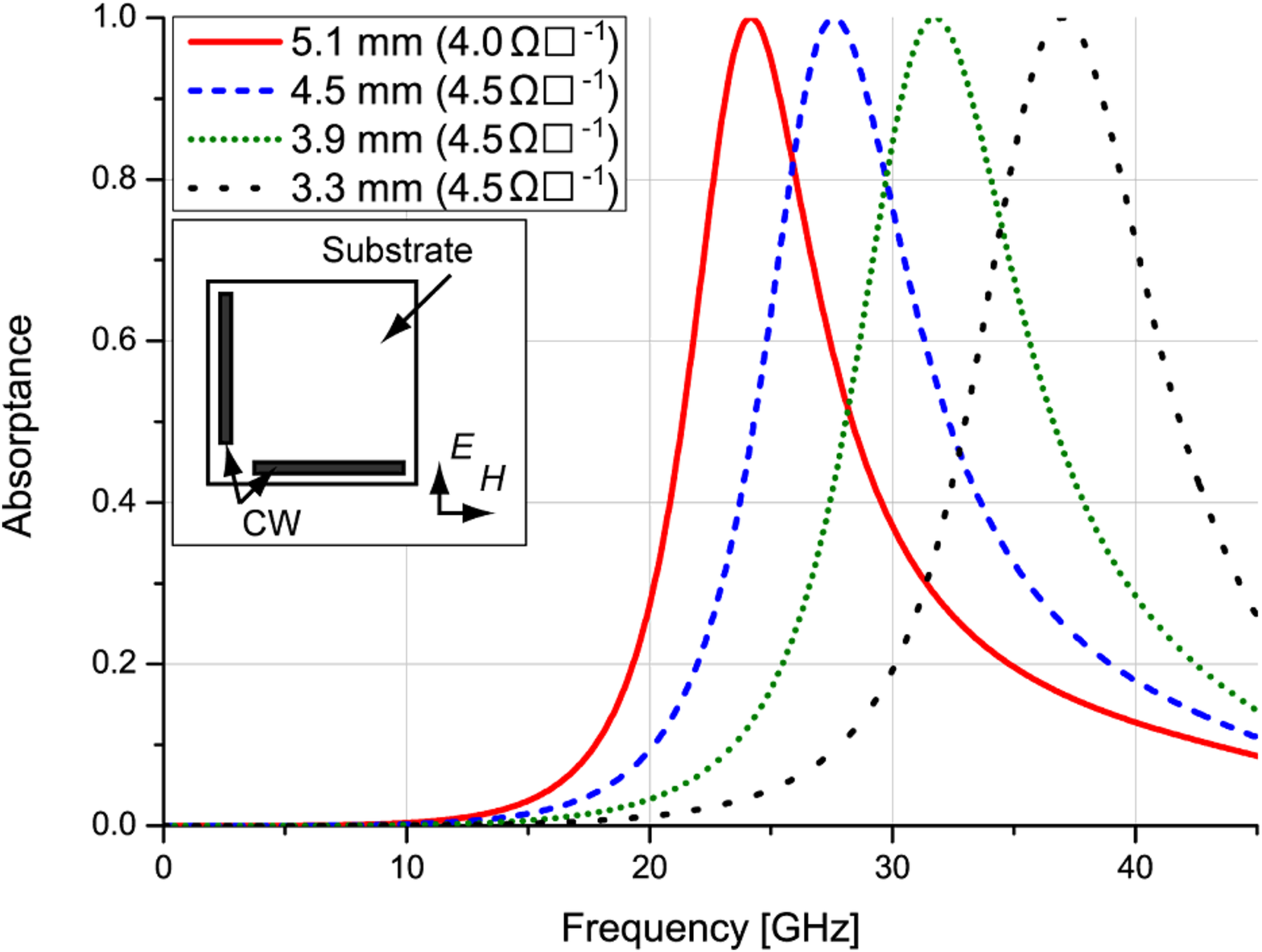}
\end{minipage}
\begin{minipage}{0.5\hsize}
\centering
\includegraphics[width=\linewidth]{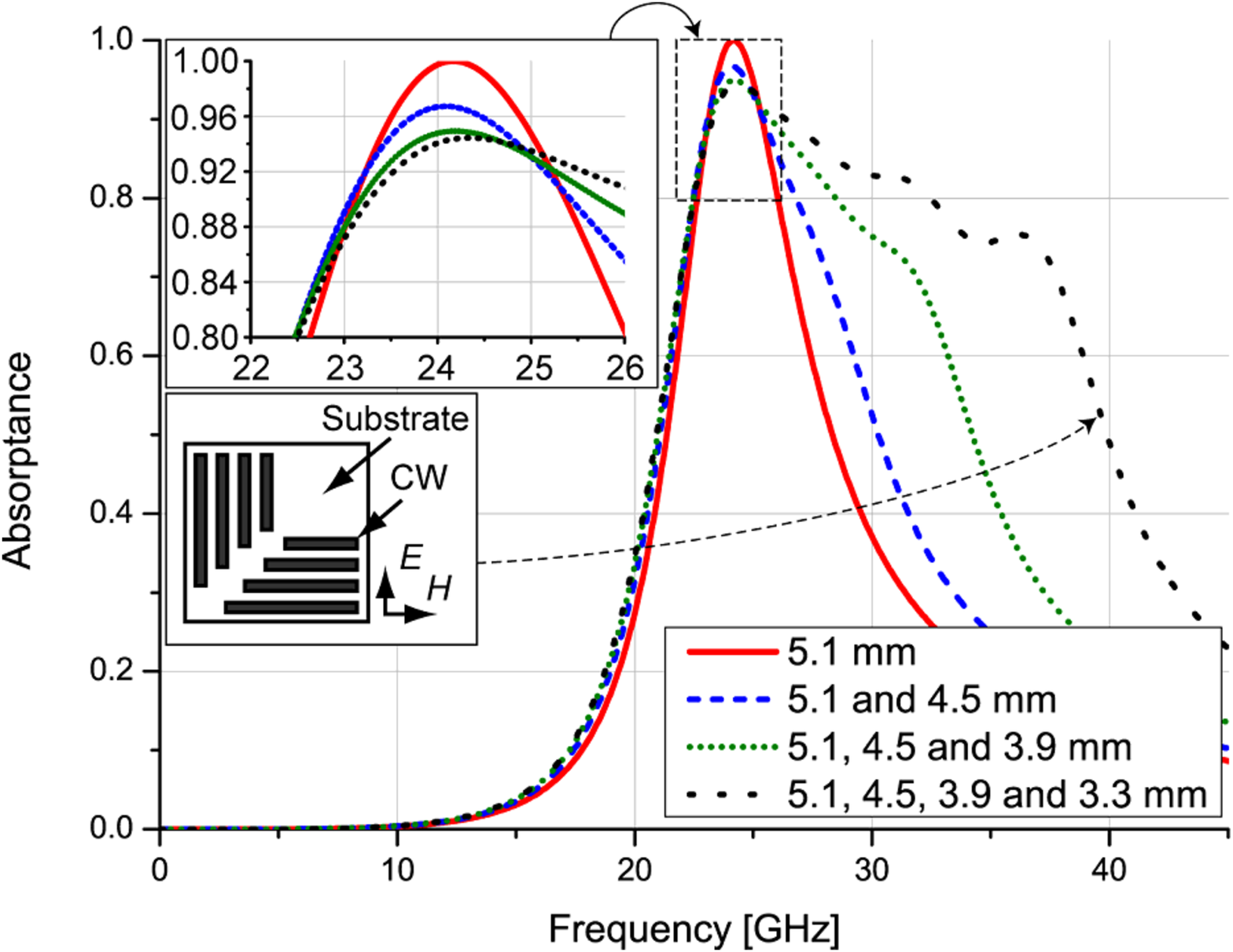}
\end{minipage}
\begin{minipage}{0.5\hsize}
\centering
\includegraphics[width=\linewidth]{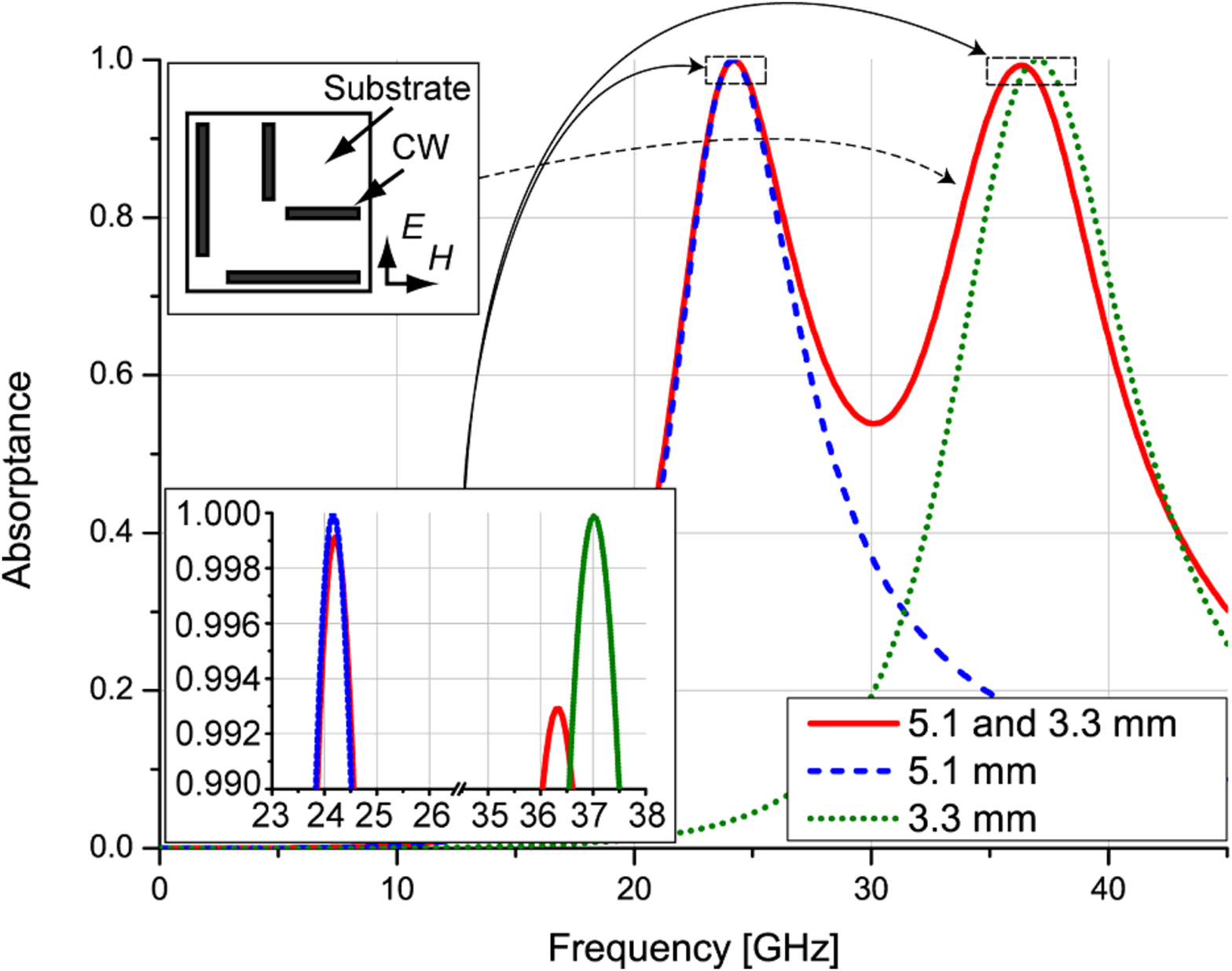}
\end{minipage}
\begin{minipage}{0.5\hsize}
\centering
\includegraphics[width=\linewidth]{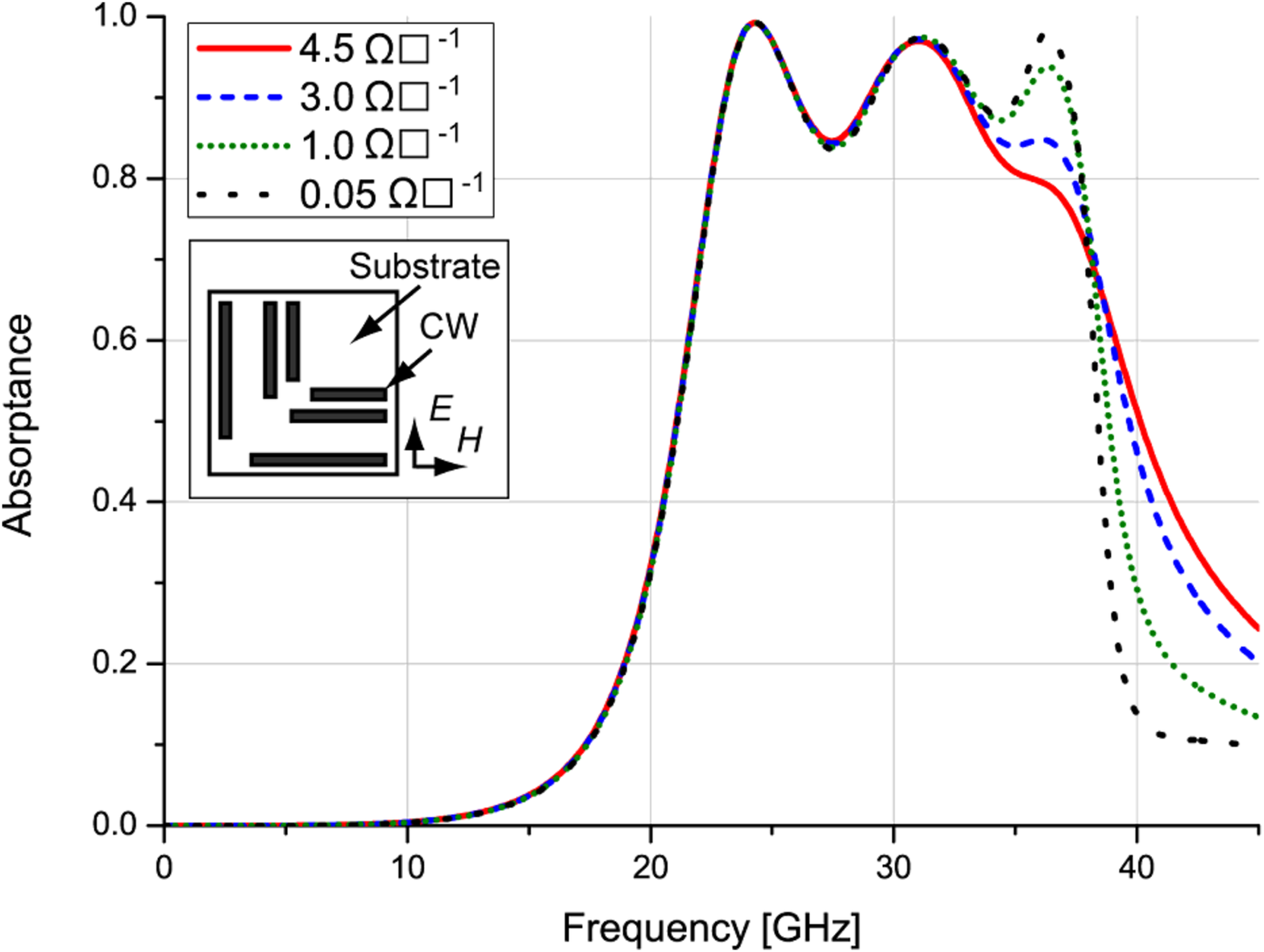}
\end{minipage}
\begin{minipage}{0.5\hsize}
\centering
\includegraphics[width=\linewidth]{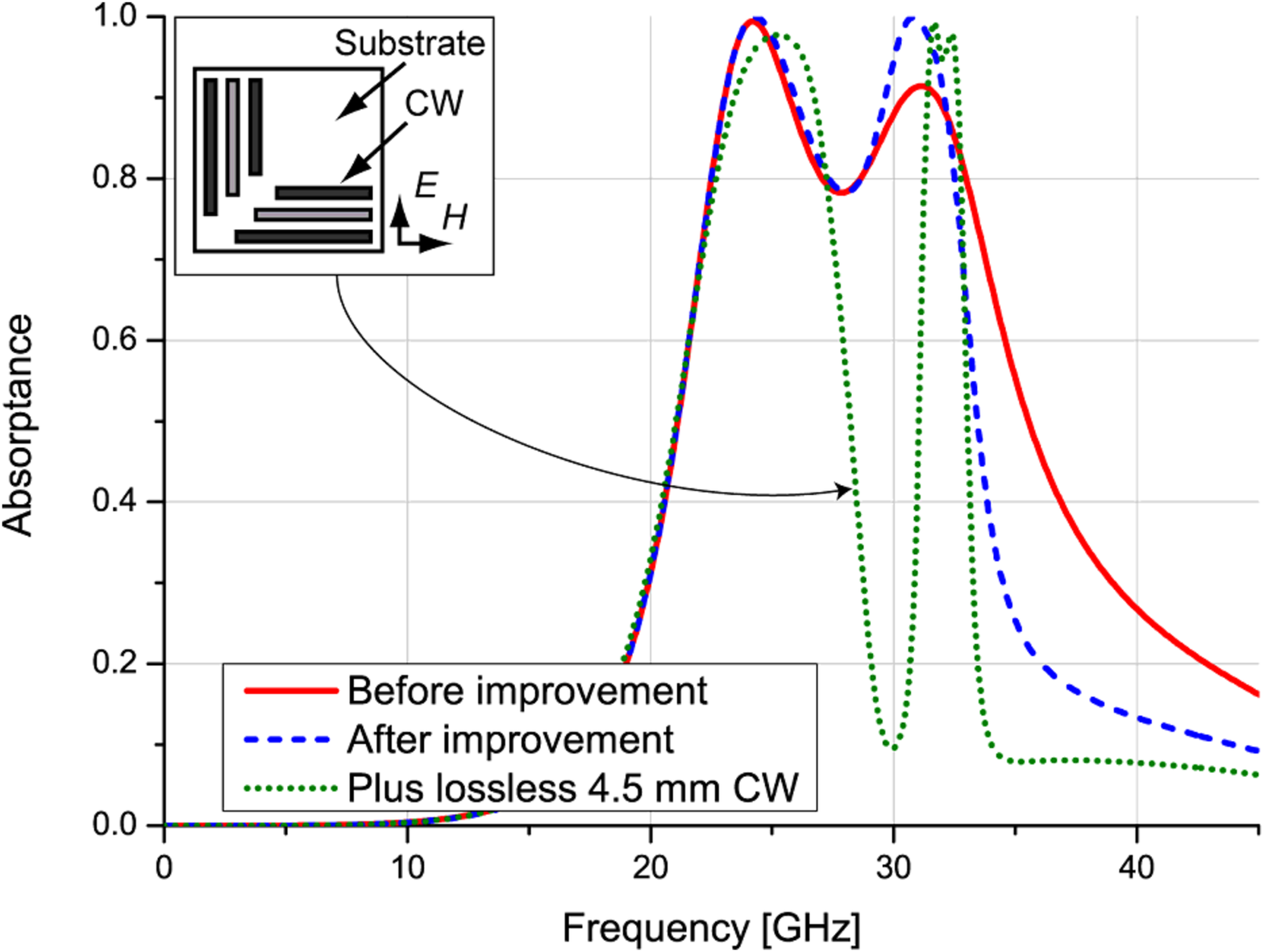}
\end{minipage}
\caption{Absorptance of single CW metamaterial placed on a PEC wall. (a) shows absorptances of different lengths of CWs. The inset describes the simulated situation. In (b) another pair of the CW used in (a) is placed orthogonally to the first, as is illustrated in the inset. In (c) part or all of the CW pairs are combined as one metamaterial unit. The inset shows the structure having all the CW pairs. (d) shows that use of two CW exhibits a double absorptance peak close to the individual peaks. (e) illustrates absorptance of three CW pairs (5.1, 3.9 and 3.3 mm). Modification of the resistance value used for the 3.3 mm CW leads to a triple absorptance peak. The resistance values used for the 3.3 mm CW are shown in the legend. The other resistance values are the same as those of (b). In (f) absorptance of two CW pairs of 5.1 and 3.9 mm is improved by use of adjusted resistance values. In addition, the use of a lossless 4.5 mm CW pair exhibits a strong reduction of the absorptance at about 30 GHz. }
\label{fig:backMetalMulti}
\end{figure}

This section introduces CW metamaterial structures which exhibit very effective absorption characteristics. In addition, it is shown that broadband absorption is achieved when different lengths of CWs are combined in one periodic metamaterial absorber unit. Strong absorptances were observed when single CW metamaterial was placed on a PEC surface, as illustrated in the inset of Fig. \ref{fig:backMetalMulti} (a). In these simulations the CW positions are the same with those used in the inset of Fig.\ \ref{fig:diffLength} (b), while the resistance values were optimised to extract the maximum absorptances. The calculation results are shown in Fig. \ref{fig:backMetalMulti} (a) and the strong absorptances of $A\sim 1.0$ are seen. 

Considering the results obtained, the case of the CW with the metal backing gives the highest absorptance ($A\sim1.0$) followed by two CWs without the metal backing ($A\sim0.75$). Finally a single CW gives the lowest absorption ($A\sim 0.5$). We note that the conduction currents in each case are different as shown in Fig.\ \ref{fig:modelField} (d). 

%
Next the same pattern of CWs was deployed along the $y$ axis to reduce polarisation dependence. The calculation results are shown in Fig. \ref{fig:backMetalMulti} (b), where the sheet resistance of each CW were again optimised. The inset of Fig. \ref{fig:backMetalMulti} (b) describes the simulated structure. As a result of the comparison with Fig. \ref{fig:backMetalMulti} (a), strong absorptance of $A\sim 1.0$ is still maintained despite the use of another orthogonal CW. According to \cite{pendryCW}, straight conductors orthogonal to the electric field do not react with the external field significantly so that the electromagnetic properties of the whole structure are not significantly affected. 

To obtain broadband behaviour some or all of the pairs of the CWs used in Fig.\ \ref{fig:backMetalMulti} (b) are combined as one metamaterial unit. The inset of Fig.\ \ref{fig:backMetalMulti} (c) illustrates the structure. The calculation results are shown in Fig.\ \ref{fig:backMetalMulti} (c). The results in this figure indicate that the absorptance properties in the high frequency region are improved by use of additional lossy pairs of CW. The addition of extra CWs introduces additional absorption peaks, as shown in Fig.\ \ref{fig:backMetalMulti} (d) where two CW pairs of 5.1 and 3.3 mm show two absorptance peaks corresponding to those of the individual CWs. Although such behaviour has been recently reported \cite{dual1,dual2}, the advantage of using the lossy CWs introduced here is to easily customise the absorptance characteristics for \emph{both} polarisations. Furthermore, as is explained below, it is also easy to realise several absorptance peaks by adding extra pairs of CW. 

The absorption characteristics may be further enhanced by optimising the sheet resistance values. This is shown in Fig.\ \ref{fig:backMetalMulti} (e) in which the three CW pairs of 5.1, 3.9 and 3.3 mm are deployed. In this figure the sheet resistance value of only the 3.3 mm CW is varied, while the values of the other CWs are the same as those of Fig.\ \ref{fig:backMetalMulti} (b). It is found from Fig.\ \ref{fig:backMetalMulti} (e) that the absorption characteristics are improved by use of adjusted resistance values, leading to a triple absorptance peak (0.993 at 24.28 GHz, 0.975 at 31.17 GHz and 0.981 at 36.29 GHz). Again, this structure will interact with both polarisations, compared with that of \cite{dual1,dual2} and is an important advantage of the lossy CW metamaterial absorber. Although only the triple absorptance peak is illustrated, further absorptance peaks are possible by using extra CW pairs. 

Use of additional pairs of CW enables us not only to increase absorption, but also to decrease it. In Fig.\ \ref{fig:backMetalMulti} (f), the absorptance of the CW metamaterial absorber composed of 5.1 and 3.9 mm CW pairs with the sheet resistance values used in Fig.\ \ref{fig:backMetalMulti} (b) is illustrated. This absorptance peaks can be enhanced by use of optimised sheet resistance values as is described in the figure. In addition to these two CW pairs, when the lossless CW pair of 3.9 mm is deployed (see the inset of Fig.\ \ref{fig:backMetalMulti} (f)), the absorptance magnitude between the two absorption peaks is markedly reduced due to the resonance of the 3.9 mm CW pair (where the sheet resistance values of 5.1 and 3.9 mm CWs were modified again). As expected, the centre of the absorptance reduction is close to the resonant frequency of the 3.9 mm CW (c.f. Fig.\ \ref{fig:backMetalMulti} (b)). This allows for a fully customised absorptance characteristic.

\section{Conclusion}
\label{sec:conc}
This paper demonstrated by numerical simulation that highly customisable broadband absorption for arbitrary polarisation is possible by use of conductively lossy CWs as metamaterial absorbers. To begin with, the basic properties of the conductively lossy CW metamaterial absorbers were investigated. A dependence of absorptance peak on the conductive loss was explained with a simple equivalent circuit. In paired CW metamaterials, absorptance peaks were improved by manipulating the two resonance frequency positions and by using independent sheet resistance values for the front and back CWs. The absorptance of the conductively lossy CW metamaterial was further improved, when single CW metamaterials were placed on a PEC wall. Moreover, when different lengths of CWs were combined as one metamaterial unit, broadband absorption was exhibited. The deployment of orthogonal pairs of CWs showed that the structure simulated here works for both polarisations. Due to the flexible absorptance characteristics, the idea of using conductively lossy CW pairs adds additional advantages to metamaterial absorbers and opens up a new area for metamaterial applications. The interesting properties of CW metamaterials may be further improved, when the structures are designed to absorb off--normal incident waves. This may be possible by using multiple CWs whose lengths and sheet resistance values are suitably optimised. 
\end{document}